\documentclass[aps,reprint,superscriptaddress,amsmath,amssymb,amsfonts,prb]{revtex4-1}

\usepackage{xcolor}
\definecolor{darkgreen}{rgb}{0,0.6,0}
\definecolor{cyan}{rgb}{0,0.7,0.8}
\usepackage[colorlinks=true,citecolor=darkgreen,urlcolor=cyan]{hyperref}

\usepackage{graphicx}
\usepackage{braket}

\newcommand{\mrm}[1]{\mathrm{#1}}

\newcommand{\mbb}[1]{\mathbb{#1}}
\newcommand{\mc}[1]{\mathcal{#1}}
\newcommand{\rcite}[1]{Ref.~\onlinecite{#1}}

\begin{document}


\title{Simulation of braiding anyons using Matrix Product States} 

\author{Babatunde M. Ayeni}
\email{babatunde.ayeni@mq.edu.au}
\affiliation{Centre for Engineered Quantum Systems, Dept. of Physics \& Astronomy, Macquarie University, NSW 2109, Australia}
\author{Sukhwinder Singh}
\affiliation{Centre for Engineered Quantum Systems, Dept. of Physics \& Astronomy, Macquarie University, NSW 2109, Australia}
\author{Robert N. C. Pfeifer}
\affiliation{Dept. of Physics \& Astronomy, Macquarie University, Sydney, NSW 2109, Australia}
\author{Gavin K. Brennen}
\affiliation{Centre for Engineered Quantum Systems, Dept. of Physics \& Astronomy, Macquarie University, NSW 2109, Australia}


\date{\today}

\pacs{Valid PACS appear here}

\begin{abstract}  
Anyons exist as point like particles in two dimensions and carry braid statistics which enable interactions that are independent of the distance between the particles. Except for a relatively few number of models which are analytically tractable, much of the physics of anyons remain still unexplored.
In this paper, we show how U(1)-symmetry can be combined with the previously proposed anyonic Matrix Product States to simulate ground states and dynamics of anyonic systems on a lattice at any rational particle number density. We provide proof of principle by studying itinerant anyons on a one dimensional chain where no natural notion of braiding arises and also on a two-leg ladder where the anyons 
hop between sites and possibly braid. We compare the result of the ground state energies of Fibonacci anyons against hardcore bosons and spinless fermions. In addition, we report the entanglement entropies of the ground states of interacting Fibonacci anyons on a fully filled two-leg ladder at different interaction strength, identifying gapped or gapless points in the parameter space. As an outlook, our approach can also prove useful in studying the time dynamics of a finite number of nonabelian anyons on a finite two-dimensional lattice.
\end{abstract}

\maketitle

\section{Introduction}
Anyons are point-like (quasi)particles which exist only in two-dimensional systems and have richer exchange statistics than bosons or fermions. One of the main interests in anyons is in their application to implementing fault-tolerant (topological) quantum computation.\citep{freedman2002,kitaev2003,Nayak2008} Anyons have also garnered a substantial theoretical interest since they are proposed to exist in systems as diverse as fractional quantum Hall systems and two-dimensional spin liquids,\citep{laughlin1983,halperin1984,fradkin1989,read1999,read2000,xia2004,Stern2007,Nayak2008,pan2008,stern2010,sanghun-an2011,clarke2013,mong2014,vaezi2014} one dimensional nanowires,\cite{kitaev2001,stanescu2013,klinovaja2014,Nadj-perge2014} and ultra-cold atoms in optical lattices.\cite{Pachos2012} Recent experiments showing evidence for Majorana edge modes (i.e.~Ising anyons) in nanowires\cite{Nadj-perge2014} might bring us closer to working with anyons in the lab, with far-reaching scientific and technological 
applications.

One dimensional chains of static SU(2)$_k$ anyons with a local antiferromagnetic Heisenberg-like interaction have been studied extensively since, for example, they are critical and realize all minimal models of conformal field theories (CFTs).\cite{Trebst2008} It is also natural to ask whether interesting states and phases appear in anyon models where the anyons are allowed to hop on a lattice and braid around one another. Braiding pairs of anyons generally transforms the anyonic state in a non-trivial way, in contrast with bosons and fermions which merely pick up a factor of $\pm 1$. For anyons, braiding is a topological interaction, with the meaning that the interaction is independent of the distance between the anyons and arises only from the inherent anyonic statistics. In Refs.~\onlinecite{Poilblanc2011a,PhysRevB.87.085106} the authors report on some phases that appear in lattice models of itinerant anyons, where the anyons---coupled by a Heisenberg interaction---are located on the sites of a lattice, 
with vacancies which allow for anyons to hop between sites but without braiding around one 
another. In Ref.~\onlinecite{Zatloukal:2014ys} the authors study the real time dynamics of a single anyon moving between the sites of a ladder lattice with static anyons pinned to the plaquettes of the ladder, which serves as a model of coherent noise in topological quantum memories, and uncovers a signature that distinguishes abelian anyons from non-Abelian anyons based on their transport properties. Noise models for medium sized topological memories based on real time {stochastic} dynamics of braiding Ising models anyons,\cite{Brell:2014ly} Fibonacci anyons,\cite{Burton} and quantum double model anyons \cite{Wootton:2014ve} have also been studied. In this paper, we describe how to simulate ground states of 1D and quasi-1D models of itinerant anyons, which may or may not involve braiding, and possibly include a Heisenberg interaction. We benchmark our method by reporting ground state energies and ground state entanglement for these models.

Large anyonic systems, like generic quantum many-body systems, are hard to simulate on a classical computer due to the exponential growth in the dimension of the state space with the number of particles. Until recently, numerical studies of anyons have primarily used exact diagonalization,\cite{Feiguin2007,Trebst2008, Trebst2008a, Poilblanc2011a,Poilblanc2011,pfeifer2012} which limits analysis to small system sizes and relies on finite-size scaling to extract properties in the thermodynamic limit. A more successful approach uses tensor networks (TNs) which describes quantum many body states using a network of low rank tensors which can be contracted together to compute relevant quantities such as ground state energy, correlations, subsystem entropy, etc. One of the simplest tensor networks is the matrix product state (MPS) which forms the basis of highly successful algorithms, namely, the density matrix renormalization group (DMRG)\cite{white1992,white1993,schollwock2011} and the time-evolving block 
decimation (TEBD),\cite{Vidal2003,white2004,daley2004} to 
simulate the ground state and dynamics of 1D and quasi-1D
quantum many-body systems. Exploiting translation invariance in TN states has allowed the study of systems directly at thermodynamic limit, circumventing the limitation on size encountered in exact diagonalization.\cite{Vidal2006}



Owing to their success for spin systems, tensor network algorithms have recently been 
adapted to simulate quantum many-body systems of anyons.\cite{Pfeifer2010a,Konig2010,pfeifer2012a} In particular, anyonic versions of the Matrix Product States (MPS), and of the TEBD and DMRG algorithms have been proposed and tested with a high degree of accuracy for anyonic chains.\cite{Singh,Pfeifer2015} Tensor network algorithms are adapted to anyons by explicitly hardwiring the constraints implied by the fusion rules of the anyon model into the tensor network ansatz. This provides two important advantages. First, {an anyonic TN representing a many-body anyonic state contains fewer complex coefficients than a {non-symmetric TN} description of the same state that does not explicitly encode the anyonic symmetry}, thus providing for computational speedup. Secondly, using an anyonic TN as an ansatz in numerical simulations guarantees that one remains in the physically relevant sector of the Hilbert state, namely, one with the desired total anyonic charge, and thus avoiding 
leakage into states that are not 
allowed by 
the physics of the system, due to numerical errors.


In this paper, we describe how to simulate the ground state of a system of itinerant anyons by means of the anyonic TEBD algorithm that additionally incorporates a U(1) symmetry corresponding to conservation of particle number density. Our construction of the combined Anyon $\times$ U(1) symmetric MPS is the first to allow for simulating these systems with an arbitrary, \emph{specified} rational particle number density (or filling fraction), and gives direct access to Hilbert space sectors enumerated by anyonic charge and particle number density. Our MPS ansatz also allows us to simulate bosons, fermions, and anyons using the same algorithm, since bosons and fermions can be treated as simple types of anyons.

Models of itinerant hardcore particles (bosons, fermions or anyons) in one dimension all have the same ground state properties since the particles do not exchange positions. However, in two or higher dimensions, there are several paths by which particles may exchange positions. Therefore, beyond 1D, ground state properties of hardcore bosons, spinless fermions and hardcore anyons should reflect the influence of their exchange statistics. We test our method using itinerant Fibonacci anyons on a chain and itinerant braiding (henceforth, simply ``braiding'') Fibonacci anyons on a two-leg ladder, and show how the ground state energies differ from those of hardcore bosons and spinless fermions. We also present results for the ground state entanglement of the Golden Ladder model comprised of Fibonacci anyons interacting by means of ferromagnetic or antiferromagnetic Heisenberg interactions.

{
Thematically, the paper is divided into two parts. The first part develops the Anyon $\times$ U(1) symmetric TN formalism, and the second part describes applications of this ansatz to the simulation of models of itinerant and braiding anyons. The Anyon $\times$ U(1) symmetric MPS combines the recently proposed anyonic MPS~\cite{Singh,Pfeifer2015} with the implementation of a U(1) symmetry in the MPS~\cite{singh2010,Singh2011} and as such our presentation contains some review of both elements separately, which serves both as a reminder of important concepts and also introduces useful terminologies that persist throughout the paper. The structure of this paper is as follows: in Sec.~II, we review the anyonic MPS. In Sec.~III, we review the implementation of a U(1)~symmetry in the MPS corresponding to conservation of particle number density, in particular showing how it can be achieved as an instance of the anyonic MPS and how an arbitrary filling fraction is realized at the level of the ansatz. In Sec.~IV we 
construct the combined MPS ansatz that incorporates both the anyonic symmetry and the U(1) symmetry. We present test models and benchmarking results in Sec.~V and some conclusions in Sec.~VI.
}

\section{Anyonic Matrix Product States\label{sec:AnyonicMPS}}
{We give a brief review of the anyonic MPS constructed in Refs. \onlinecite{Singh} and \onlinecite{Pfeifer2015}. For more details, the reader can consult these articles}. 

{The basis of the Hilbert space of anyons is described by a labeled directed fusion tree (see Fig.~\ref{FusionTree}) where the charge $c$ on any incoming edge at a vertex is determined from the charges $a$ and $b$ of the two outgoing edges around the same vertex, 
\begin{figure}
 \includegraphics[width=\columnwidth]{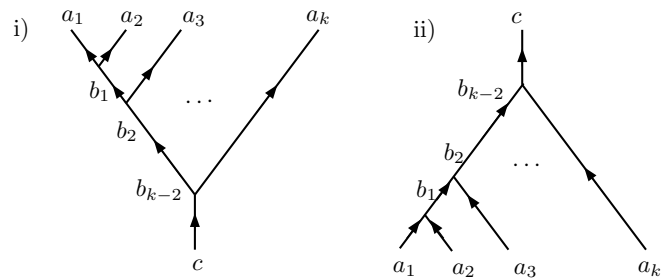}
 \caption{(i) Splitting tree and (ii) fusion tree, defining the ``ket' and ``bra'' bases respectively for a total number $k$ of anyonic charges $(a_1, a_2, \ldots, a_k)$ on the leaves, and $(b_1, b_2, \ldots, b_{k-2})$ as the fusion products on the links of the trees. The charge $c$ on the trunk can, in principle, take all possible charge values permissible by the anyon model. If however, the fusion tree defines the basis of a pure quantum state, the charge $c$ can only be the vacuum charge $\mathbb{I}$.}
 \label{FusionTree}
\end{figure}
according to the fusion rules of the anyon theory 
\begin{equation}
 a \times b \rightarrow \sum_c N_{ab}^c ~ c ,
\end{equation}
which implies that charges $a$ and $b$ are allowed to fuse to, possibly, several different charges $c$. The $N_{ab}^c$ is the multiplicity tensor, which encodes the number of ways of obtaining charge $c$ from charges $a$ and $b$. We consider only multiplicity-free anyon models in this work, with $N_{ab}^c = 0,1$, which includes some of the models most relevant to current experiment such as Ising anyons and Fibonacci anyons. When $\sum_c N_{ab}^c > 1$, the anyon model is non-Abelian.}
{Anyonic charges have \emph{quantum dimensions} analogous to the dimension of an irrep for a group, and the dimensions $d_a$, $d_b$, and $d_c$ of three charges $a$, $b$, and $c$ must satisfy
\begin{equation}
 d_a ~ d_b = \sum_c N_{ab}^c d_c,
\end{equation}
though in contrast with group theory, there is no requirement that the quantum dimensions be integer.}
{The total quantum dimension is then defined as $ \mathcal{D} = \sqrt{\sum_a d_a} $} {summing over} {all anyon charges $a$ of the theory.}

{The labeled fusion/splitting tree in Fig.\ref{FusionTree} contains many charge labels, and can be extremely verbose when dealing with large anyonic systems. While explicit labeling of fusion trees is possible in principle, it is not very practical for anyonic tensor network simulations. A better alternative {is} to enumerate the labeled fusion trees {having} a particular charge $c$ at the trunk of the tree. To this end, let $c$ is the  total charge at the trunk of the fusion tree {and} introduce a new index $\mu_c$ that enumerates each unique labeled fusion tree in increasing numerical order, $\mu_c = 1, 2, \cdots, \nu_c${. H}ere $\mu_c$ is called the degeneracy index, and $\nu_c$ is the degeneracy of the charge sector $c$.\footnote{The term ``degeneracy'' in symmetric TN does not refer to the ``degeneracy of energy levels'' as used in many-body physics, but to the number of configuration {states forming a basis} 
in a particular symmetry sector.}  All the fusion trees are therefore concisely labeled by the multi-index $\gamma=(c, \mu_c)$, with $c$ as the total charge label and $\mu_c$ as its degeneracy index. A tensor network consists of connected tensors, which may be a combination of, single-index tensors (or vectors), two-index tensors (or matrices), or multi-index tensors. Similarly, the tensor objects of our Anyonic-U(1) MPS ansatz are the anyonic analog of the non-symmetric tensors, although, for anyons, we do not permit using tensors with more than three legs, as the {fusion tree labelling of non-Abelian tensors} can no longer be uniquely specified {purely by multi-indices on the external legs}. The details of how to construct anyonic tensors are given in the Appendix~\ref{AnyonicTensorNetworks}}.

One convenient form of the conventional MPS ansatz is that given by Vidal,\cite{Vidal2003} which is an array of two-index and three-index tensors forming a linear network of tensors. For a finite lattice with open boundary condition, the tensors on the boundary of the MPS (i.e.~the first and last sites) are two-index tensors while the ``bulk'' of the network consists both of two-index tensors (Schmidt vectors) and three-index tensors for each of the other $(n-2)$ sites. 

Analogously, the MPS was adapted to anyons by Singh et.~al. in Ref.~\onlinecite{Singh}, using the basic anyonic tensors (two-index and three-index anyonic tensors) after the pattern of the conventional MPS. Each three-index tensor is indexed by both the charge and the degeneracy of the anyons making up each site. The charges on the trivalent vertex of the tensor are compatible in accordance with the fusion rules of the anyon model. The Schmidt vectors, which are two-index tensors, are charge-conserving diagonal matrices. The basis labeling $\alpha_i=(a_i, \mu_{a_i})$ for each site of the anyonic lattice is given by the set of charges $a_i$ and the degeneracies $\mu_{a_i}$ of each charge. The labels $\mu_{a_i}$ take fixed value 1 if there is only one configuration for each possible charge labeling at each site, e.g.~if the possible physical states are merely the presence or the absence of a charge.

Formally, for a lattice $\mathcal{L}$ of $L$ sites with anyonic charges $\alpha_1=(a_1, \mu_{a_1}), \alpha_2=(a_2, \mu_{a_2}), \ldots, \alpha_L=(a_L, \mu_{a_L})$, the anyonic MPS encoding the ground state $\Psi_{\text{GS}}$ is given diagrammatically as 
\begin{equation}
 \includegraphics[scale=0.9]{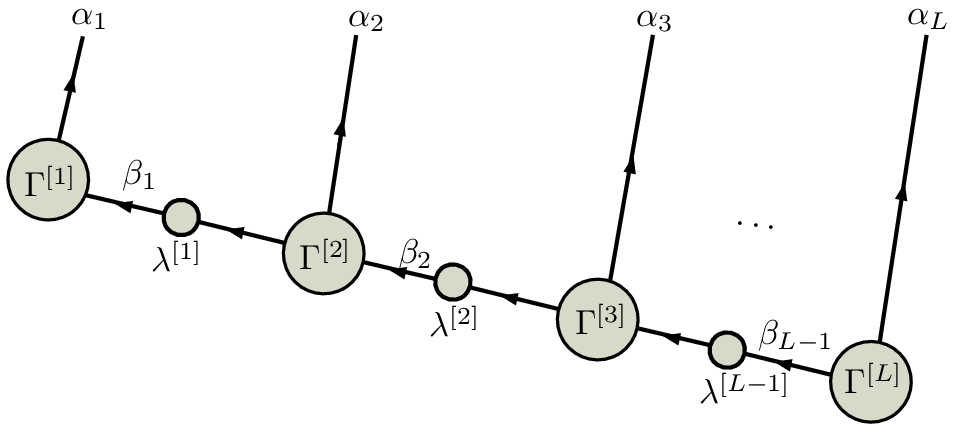}
\label{fig:AnyonicMPS}
\end{equation}
where the multi-indices $\beta_{i+1} = (b_{i+1}, \mu_{b_{i+1}})$ on the bonds {are obtained by an iterative fusion of the multi-indices $\beta_{i} = (b_i, \mu_{b_i})$ and $\alpha_{i+1}=(a_{i+1}, \mu_{a_{i+1}})$, 
\begin{equation}
 \beta_i \times \alpha_{i+1} \rightarrow \beta_{i+1}, 
\end{equation}
where, as before, the charge $b_{i+1}$, 
\begin{equation}
 b_{i+1} = \sum_{b_i, a_{i+1}} N_{b_i a_{i+1}}^{b_{i+1}} (b_i \times a_{i+1}), 
\end{equation}
and the total degeneracy $\nu_{b_{i+1}}$ of the charge $b_{i+1}$ is determined by 
\begin{equation}
 \nu_{b_{i+1}} = \sum_{b_i, a_{i+1}} N_{b_i a_{i+1}}^{b_{i+1}} \nu_{b_i} \nu_{a_{i+1}}. 
\end{equation}
}

{It should be noted that} this anyonic MPS has been drawn with site indices going upwards, to make apparent the visual similarity with anyonic fusion tree diagrams, but it is essentially the same ansatz given in Ref.~\onlinecite{Singh}. Due to the iterative fusion process down the tree of the anyonic MPS {the dimensions of the tensors $\Gamma^{[i]}$ required to exactly construct an arbitrary state will}
vary, but {in practice} {an upper bound is} imposed on the bond dimension $\chi$ ahead of time. The {bound} chosen usually depends on the amount of entanglement and correlations needed to faithfully represent the state of the system (and on computational resources available). As such, anyonic MPS provides a systematic way of handling anyonic systems, specifying both the basis (i.e.~the fusion tree) and encoding the amplitudes of the state in the tensors. 

{As {a proof-of-principle} example, this anyonic-MPS ansatz has been {used} to simulate, together with {the} anyonic-TEBD algorithm, a chain of interacting non-Abelian anyons (e.g. Fibonacci and Ising anyons) coupled by a Heisenberg interaction. The charge multi-index $\alpha_i$ on each site $i$ of the leaves of the anyonic-MPS is set (in the case of Fibonacci anyons) {to} $\alpha_i = (\tau, 1)$, where $\tau$ is the Fibonacci anyon charge, and the number $1$ is the degeneracy of the $\tau$ charge on site $i$ (i.e. the number of {different configurations} on the site {consistent with a total charge of $\tau$}). {T}he anyonic MPS is{, however, a} general {ansatz capable of} dealing with systems with any quantum group symmetry, and hence, can be adapted to work with other symmetries, Abelian or non-Abelian. For instance, by replacing the anyonic charges 
with particle number charges, the anyonic-MPS can serve as a U(1)-MPS~\cite{Singh2011}, which can be used to simulate physical systems having a global particle number $N$ on a finite lattice $\mathcal{L}$.} 

{On an infinite lattice {with} translation invariance of the Hamiltonian, {if the U(1) charge is identified with particle number then} the U(1)-MPS is primitively a zero-density ansatz {[i.e. one favouring a mean U(1) charge per site of 0],} and cannot directly be used to simulate an infinite lattice with a finite non-zero {particle} density. In the next section we show how to tune the U(1)-MPS to simulate an infinite lattice system at non-zero density, and in Section \ref{Anyon-U1MPS} we propose a modified ansatz, the Anyonic-U(1) MPS, that conserves both particle density and anyonic charge symmetry, and which can be used to simulate anyonic systems (including braiding of anyons) at a specified rational {filling fraction}.}

\section{U(1)-MPS and Particle Density Conservation} \label{U1-MPS}
In the last section we alluded to the fact that the anyonic MPS can serve as a U(1)-MPS by replacing the anyonic charge labels with the particle number charge labels. Specifically, let us consider a lattice $\mathcal{L}$ of $L$ sites, where each site can accommodate a finite number of particles, $n=0,1,2, \ldots, d-1$. The positive integers $n$ can be regarded as the irreps of the $U(1)$ symmetry, which can intuitively be understood as: $n=0$ is the absence of a particle, $n=1$ is the presence of one particle, $n=2$ is the presence of two particles, and so on. The total number of particles $N$ on the lattice of $L$ sites is $N=\sum_{i=1}^L n_i$, {with} a {particle} density of $\nu=N/L$.

The Hilbert space of the lattice, $\mathbb{V}^{\mathcal{L}} = \bigotimes_{i=1}^L \mathbb{V}^{(i)}$, can be alternative written as,  $\mathbb{V}^{\mathcal{L}} = \bigoplus_{n=0}^N \mathbb{V}_n$, {a} direct sum {over} subspace{s} with fixed number{s} of particles $n$. Utilizing this alternative structure a particle-number conserving Hamiltonian $\hat{H}$ can be directly diagonalized in the $\mathbb{V}_n$ subspace, offering savings on the computational cost. The U(1)-MPS ansatz for $N$ particles on an $L$-site lattice can be derived from the anyonic MPS by fixing the particle number $N$ and degeneracy $\nu_N=1$ at the ``right end'' of the last tensor, and charge $0$ (i.e. zero) on the ``left end'' of the first tensor. The on-site multi-indices of the ``bulk'' $(L-2)$ tensors carry $\alpha_i = (n_i, \mu_i)$, where $n_i$ is the U(1) charge on site $i$, and $\mu_i$ enumerates the degeneracy of that charge, for all $i \in \mathcal{L}$. 
The MPS bonds also carry charge and degeneracy 
indices, but unlike systems of anyons where degeneracy comes from the fusion rules of the anyon model, degeneracy in U(1)-symmetric lattice models comes from the number of combinatorial arrangement of the charges on the lattice.

Therefore, with a properly constructed ansatz and an optimization algorithm like TEBD or DMRG, one can compute the ground state of a local U(1)-symmetric Hamiltonian {on} a finite lattice. If this finite U(1)-MPS is naively extended to simulate an infinite lattice model, the ansatz would correspond to a zero-density ansatz because of the finite size of the bond dimension $\chi$ {and the assumption that the U(1) charge labels exhibited on this bond are finite}. In the next subsections we give a heuristic proof of this statement, and we then propose a {technique} which can be employed to tune the U(1)-MPS away from being a zero-density ansatz, to any desired non-zero {particle} density.

\subsection{Zero-density U(1)-MPS} \label{Zero-Density-U1-MPS}
\begin{figure}
 \includegraphics{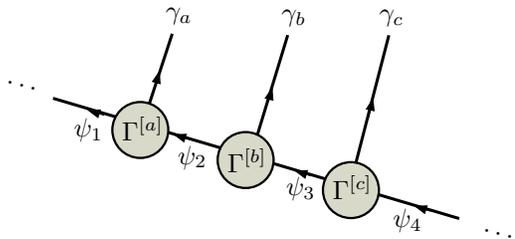}
\caption{An example of an infinite MPS with a block made up of three tensors $\Gamma^{[a]}$, $\Gamma^{[b]}$ and $\Gamma^{[c]}$. By translational invariance of the Hamiltonian, an infinite MPS corresponds to an infinite repetition of the block and hence optimization to ground state is performed only on the tensors within a single block. }
\label{InfiniteAnyonicMPS}
\end{figure}

Restricted to a finite bond dimension $\chi$ carrying finite U(1) charges, the U(1)-MPS with integer charge labels on an infinite lattice is a zero-density MPS ansatz. Consider a section of the infinite MPS in Fig.~\ref{InfiniteAnyonicMPS} with the charge-degeneracy indices on physical sites
$$ 
\alpha_1 = (n_1, \mu_{n_1}), \alpha_2 = (n_2, \mu_{n_2}), \alpha_3 = (n_3, \mu_{n_3}) ,
$$ 
and on the links 
\begin{align*}
\beta_1 &= (m_1, \mu_{m_1}), \beta_2=(m_2, \mu_{m_2}), \beta_3=(m_3, \mu_{m_3}), \\
& \beta_4=(m_4, \mu_{m_4}) .
\end{align*}

The on-site charges $n_i$ are set to take positive integer charges corresponding to particle number (e.g.~hardcore boson has $n_i{\in\{0,1\}}$). The charges $m_i$ on the links take only a finite number of charges with degeneracy index $\mu_{m_i} = 1, 2, \cdots, \nu_{m_i}$. The charges and degenerac{ies} on the bond are constrained by the finite bond dimension $\chi$ and given as $\chi=\sum_{m_i} \nu_{m_i}$, where $m_i$ labels the charge on the link $i$. For any realistic computer simulation, the charge labels on the MPS bonds are all finite.
Assume we cut the infinite lattice into two partitions. There exists a finite amount of charge $k$ on the link of the left partition, corresponding to a finite number of particles, and the density on the left half-chain is therefore $\nu=k/\infty \rightarrow 0$ and therefore the infinite U(1) MPS is a zero{-}density ansatz. However it is possible to remedy this and have a nonzero density U(1) MPS by shifting the on-site charges {so that a U(1) charge of zero corresponds to} the desired filling fraction. We present this transformation below.

\subsection{Non-zero density U(1)-MPS}
{By employing translation invariance, an infinite U(1){-symmetric} MPS consists of a block of repeated U(1)-symmetric tensors, albeit that {such an} ansatz is zero-density and will yield a ground state of an empty lattice {as seen above.} However, by transforming the on-site charges of the MPS, we can {cause a U(1) charge of zero to correspond to} the desired density. }

{For simplicity and without loss of generality we consider hardcore particles, with charge labels $n{\in\{0,1\}}$ on each site of the U(1)-MPS lattice. Let the desired density on the infinite lattice be $\nu=p/q$, which can be interpreted as having {an average of} $p$ particles on {every} $q$ sites. Using the additive (abelian) fusion rules of U(1) charges, a U(1)-MPS with $p$ particles corresponds to having $p$ sites with charge $n=1$ and the remaining $q-p$ sites with holes $n=0$. In an infinitely increasing block, the number of particles $p$ increases infinitely, but by ``subtracting off'' the $p$ number of particles, we can re-center the relevant subspace to be labeled by the charge $0$, which is retained in a practical simulation. Formally, by using the transformation, 
\begin{equation}
 n' = q\left(n-\frac{p}{q}\right)=qn - p, \quad p \leq q,\label{eq:U1rescale}
\end{equation}
the on-site charges transform as,
\begin{align}
 n=0 \quad \rightarrow \quad n' = -p \nonumber \\
 \quad \quad n=1 \quad \rightarrow \quad n'=q-p \nonumber
\end{align}
{where multiplication by $q$ in Eq.~(\ref{eq:U1rescale}) is purely for convenience and ensures that the $n'$ charges, like the $n$ charges, are integer.}
In essence, before this transform, the desired filling fraction in the MPS would correspond to having $p$ occurrences of charge $1$ and $q-p$ holes $0$, summing to a total charge of $p$. But after the transform, the desired filling corresponds to having $p$ occurrences of particles with charge $q-p$, and $q-p$ holes with charge $-p$, which sums to a total charge of zero. The charge distribution on any link on a U(1)-symmetric infinite MPS is centered on the zero charge sector{, which now corresponds to a particle density of {$p/q$}}. Thus it becomes possible to tune the U(1)-MPS to the desired filling {fraction} without using tensors with more than three legs.\footnote{Alternative approaches for identifying a U(1) charge of zero with the desired filling fraction, for example by inserting ancillary indices which remove $p$ U(1) charges every $q$th tensor, either add extra tensors to the network or require tensors with more than three legs, both of which are 
undesireable as they increase the complexity of the network.}
An example of how this transform applies to the half-filling is presented as an example below.} 

\subsubsection{Example: Half-filled MPS ansatz for hardcore bosons}
\begin{figure}
 \includegraphics{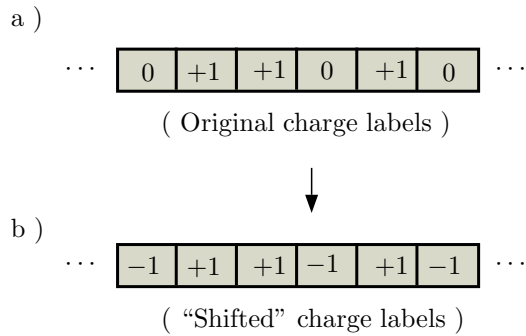}
\caption{Schematic representation of an infinite lattice with a) a typical half-filled configuration with one particle on every two sites, and b) a ``shifted'' version of a) but with average of zero particle on every two sites.}
\label{HalfFilling}
\end{figure}

{Consider a particular configuration of an infinite lattice at half filling, where there is on average, one particle on every two sites as shown in Fig.~\ref{HalfFilling}(a). Each box represents a site and the charge on the site is indicated inside the box. There is {on average} one particle {for} every two sites, {and assuming that this average density is maintained, this will} correspond to half {filling on the} infinite lattice. {T}his is {of course} not the only way to achieve half filling, but {the example will} suffice to illustrate how to achieve a half-fill{ed} U(1)-MPS. }

{With only nonnegative charges on each site, i.e. $n{\in\{0,1\}}$, the charges on the links of the MPS---which are derived by fusion of all charges leading to that link---are also all nonnegative. However the implementation requirement that the charge indices be finite (and the finite size of the bond dimension) places an upper bound on the set of charges on the links which are retained after truncation of the Hilbert space of the link. Hence the dominant larger-$N$ states in the infinite lattice are truncated. However, by using $n'=2n-1$, the on-site charges are re-defined as $0 \rightarrow -1$ and $1 \rightarrow 1$, to give the ``shifted'' configuration in Fig.~\ref{HalfFilling}(b), {for which} the dominant states now {inhabit} the zero particle sector. Nearby charge sectors such as $\pm 1$ on the bonds represent small fluctuations in filling {fraction} relative to {a baseline of $\nu=1/2$}. We emphasize that the 
complex amplitudes of the state are 
not changed, {only that} their index {is} relabeled. }

\section{Anyon $\times$ U(1)-symmetric MPS}\label{Anyon-U1MPS}
\subsection{Composite charges and fusion rules}
In the last Section, we reviewed the U(1) MPS and explained how to achieve an arbitrary rational filling fraction on the infinite lattice. 
In this Section, we investigate how anyonic systems at arbitrary filling fractions can be simulated using an ansatz that conserves both the anyonic (quantum group) symmetry and the U(1) symmetry. 

We first recognize that the two symmetry groups are described by particle spectra with differing fusion rules. Similar to creating a new group from product of two groups, we introduce the Cartesian product of the anyonic charge spectrum $\mathcal{A} = \{ a,b, c, \cdots, d \}$ and the U(1) charge spectrum which will be designated as $\mathcal{U}= \{ n,m, \cdots, z \}$ where $n$ are integer charges, $n \in \mathbb{Z}_{\infty}$. The product of the two particle spectra is given as $\mathcal{A} \times \mathcal{U} = \{ (a,n) \quad | ~ a \in \mathcal{A}, n\in \mathcal{U} \}$, where the label $(a,n)$ is referred to as the composite charge. The charges on the physical site and on the links of the MPS are taken from this set $\mathcal{A} \times \mathcal{U}$. 

The ``new'' fusion rules for the composite charges are derived from the fusion rules of the two theories,
\begin{align}
 (a_1, n_1) \times (a_, n_2) & = (a_1 \times a_2, n_1 \times n_2)  \nonumber \\
& = \sum_{a_{12}} \left(N_{a_1 a_2}^{a_{12}} a_{12},  n_1 + n_2 \right) ,
\end{align}
where as aforementioned $n_1 \times n_2$ has a unique outcome $(n_1 + n_2)$ with an additive fusion rule, while the nonabelian anyons have generally more than one fusion outcome, hence the need for the summation $\sum_{a_{12}}$ over all possible charge outcomes $a_{12}$. 

\begin{figure}
 \includegraphics[scale=0.8]{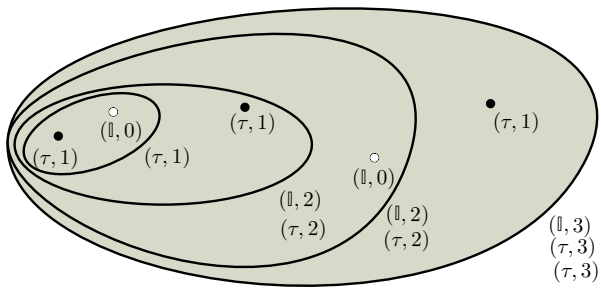}
\caption{Fusion of composite charges situated on a manifold which supports either a single nontrivial anyonic Fibonacci charge $(\tau, 1)$ or a vacuum charges $(\mathbb{I},0)$ at each point. A lattice may then be embedded into the manifold. In the figure, a linear ordering has been defined and fusion proceeds from left to right. The total charge outcomes which are $((\mathbb{I},3), (\tau,3), (\tau,3))$, indicates that there are in total three Fibonacci charge $\tau$ on the manifold fusing either into the vacuum charge ($\mathbb{I}$) or the Fibonacci anyon $\tau$ channel. As there are two charges with label $(\tau,3)$, we would also introduce a degeneracy index $\mu_{(\tau,3)} = 1 ,2$ to enumerate these outcomes. }
\label{FusionIsland}
\end{figure}

We consider only hardcore anyonic particles, meaning that either there is a nontrivial anyonic particle on a site or the site is vacant. The vacuum charge of the composite charge spectrum $\mathcal{A} \times \mathcal{U}$ is $(\mathbb{I}, 0)$. The presence of a single nontrivial anyonic charge is represented by $(a, 1)$ where $a \in \mathcal{A} \setminus {\mathbb{I}}$ and the U(1) charge $1$ imposes a hardcore constraint of a single charge on the site. The use of the U(1) charge allows the counting of the anyonic charges fusing into a particular fusion channel irrespective of the outcome anyonic charge. A simple example is shown in Fig.~\ref{FusionIsland}. 

The anyonic MPS ansatz and the U(1) symmetry discussed in previous Sections can be used together to realize an Anyon $\times$ U(1)-symmetric MPS ansatz with the desired particle density. The minor modification needed in the new ansatz involves using the composite charges along with the composite fusion rules. To have an ansatz for a particular anyonic filling fraction, the method of shifting the U(1) charges can be employed. This only amounts to a shift in the U(1) charge labels, while the labels on the anyonic fusion space are not altered. The diagrammatic representation of tensors with the new symmetry group and the MPS ansatz constructed from them are the same as given in Section~\ref{AnyonicTensorNetworks} and we will not reproduce them here.

\subsection{Manipulations of Anyon $\times$ U(1) tensors}
Topological manipulations such as F-moves, R-moves, and vertical bends applied to anyonic fusion trees are also modified in the case of a $\text{Anyon}\times \text{U(1)}$ symmetry. Let the label $\tilde{a} = (a, n)$ be the composite charge where $a$ is the anyonic charge and $n$ is the U(1) charge. Below we present typical manipulations needed to contract anyonic tensors during optimization of anyonic MPS.

\subsubsection{F-moves }
The first topological manipulation required is that of changing the fusion order of the composite charges represented by the fusion tree. Let the basis fusion tree where fusion of charges proceeds from left to right be referred to as the standard basis. If instead a different fusion ordering is chosen, such as fusion from right to left, the charge outcomes are still the same, a fact guaranteed by the constraint of associativity. Formally, this associativity constraint corresponds to the Pentagon Equations, as given in e.g.~\rcite{Bonderson2007}. The corresponding operation if the F-move, which transforms from one fusion basis to another one and is given diagrammatically as 
\begin{equation}
 \includegraphics{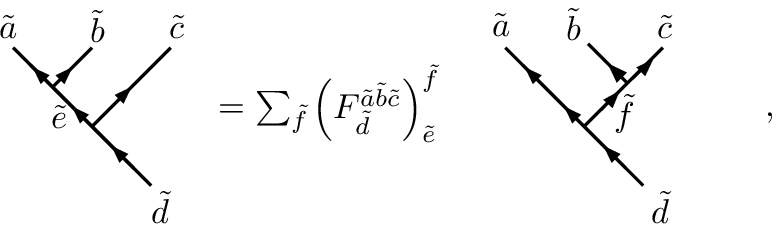}
\end{equation}
where the coefficient $\left( F^{\tilde{a} \tilde{b} \tilde{c}}_{\tilde{d}} \right)_{\tilde{e}}^{\tilde{f}} $ decomposes into its anyonic and U(1) counterparts as,
\begin{equation}
 \left( F^{\tilde{a} \tilde{b} \tilde{c}}_{\tilde{d}} \right)_{\tilde{e}}^{\tilde{f}} = \left(F^{abc}_d \right)_e^f ~ \left(F^{n_a n_b n_c}_{n_d} \right)_{n_e}^{n_f} .
\end{equation}
The factor $\left(F^{abc}_d \right)_e^f$ is given by the F coefficients of the anyon model while the U(1) factor is given by $\left(F^{n_a n_b n_c}_{n_d} \right)_{n_e}^{n_f} = N_{n_a n_b}^{n_e} N_{n_e n_c}^{n_d} N_{n_b n_c}^{n_f} N_{n_a n_f}^{n_d}$ which equals one if the charges are compatible or zero otherwise. F-moves may also be applied to pairs of contiguous vertices appearing within a larger diagram. 

It was noted in Ref.~\onlinecite{Singh2012} that a symmetric tensor decomposes into a linear superposition of the degeneracy tensor and its spin network for systems with nontrivial symmetries such as SU(2), and more generally also for quantum symmetries. Therefore any section of the anyonic MPS can be decomposed into its degeneracy tensor and anyonic network as
\begin{equation*}
 \includegraphics[width=\columnwidth]{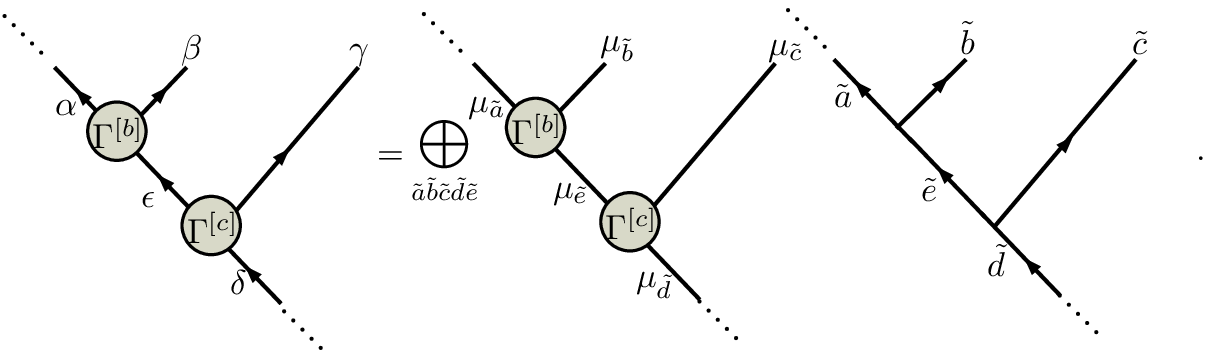}
\end{equation*}
The F-move is then applied on the anyonic diagram and the resulting F-factors are absorbed into the tensor resulting from contraction of the degeneracy tensor network. As shown, this process is valid for any portion of the diagram where the F-move operation can be applied.

\subsubsection{R-moves}
Anyons have very rich particle exchange statistics which are neither bosonic nor fermionic. The exchange factors are encoded in the R-matrix which is a matrix representation of the braid (or R-) move. The braid operator for composite anyonic charges is given diagrammatically as
\begin{equation}
 \includegraphics{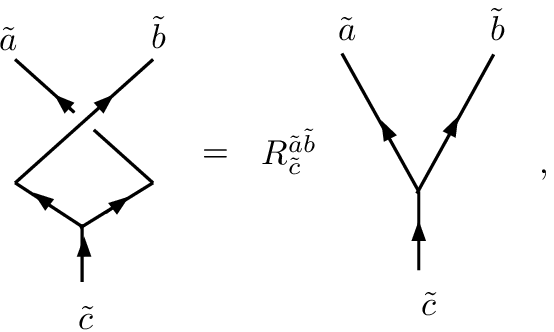}
\end{equation}
where the factor $R_{\tilde{c}}^{\tilde{a} \tilde{b}}$ decomposes as
\begin{equation*}
 R_{\tilde{c}}^{\tilde{a}\tilde{b}} = R_c^{a b} R_{n_c}^{n_a n_b} ,
\end{equation*}
and $R_{n_c}^{n_a n_b} = 1 $ if $n_a + n_b = n_c$.  The factors $R_c^{ab}$ are given by the anyon model.

To model the braiding of anyons by exchanging the positions of anyons, the Hamiltonian $\hat{H}$ should contain the braid operator. Later, we construct a Hamiltonian for the braiding of anyons supported on the vertices of a ladder.

\subsubsection{Fusion tensor and loop factors}
A trivalent tensor can be used to define a linear map from the tensor product of two Hilbert spaces $\mathbb{V}^{(A)}$ and $\mathbb{V}^{(B)}$ (which can possibly be degenerate) to a new composite space $\mathbb{V}^{(C)}$. The dimension of the new space $\text{dim}(\mathbb{V}^{(C)}) = \text{dim}(\mathbb{V}^{(A)}) \times \text{dim}(\mathbb{V}^{(B)})$. The linear map can be written as
\begin{equation}
T = \sum_{a,b,c} T_{ab}^c ~ \ket{c} \bra{a} \otimes \bra{b} ,
\end{equation} 
which sends a product basis $\ket{a} \otimes \ket{b}$ to the basis state $\ket{c}$. This assignation can be manually defined by, for instance, iterating slowly over the basis labeled by $a$ and fast over the basis labeled by $b$, sending them to a unique new basis indexed by $c$. The coefficients of $T_{ab}^c$ are 1 for a valid map $(ab \rightarrow c)$ and zero when there is no valid map. Consider the following example. Let $\mathbb{V}^{(A)}$ be a vector space of dimension $d_A$: $\mathbb{V}^{(A)}={\rm span}_{\mathbb{C}}\{\ket{x}_A\}_{x=0}^{d_A-1}$ and similarly let $\mathbb{V}^{(B)}$ be a vector space of dimension $d_B$: $\mathbb{V}^{(B)}={\rm span}_{\mathbb{C}}\{\ket{y}_B\}_{y=0}^{d_B-1}$. The tensor product space $\mathbb{V}^{(C)}= \mathbb{V}^{(A)}\otimes \mathbb{V}^{(B)}$ can be labeled by states $\{\ket{z}_C\}$ using the simple assignment map $\ket{x}_A\otimes \ket{y}_B\rightarrow \ket{z=d_B x + y}_C$. 

\begin{table}
\begin{ruledtabular}
\begin{tabular}{  c  c  c  c  c  c  }
a  &  $\mu_a$  &  b   &  $\mu_b$  &   c   & $\mu_c$   \\ \hline
0  &    1    &  0   &     1   &   0   &   1            \\ 
0  &    1    &  1   &     1   &   1   &   1             \\  
1  &    1    &  0   &     1   &   1   &   2              \\ 
1  &    1    &  1   &     1   &   0   &   2               \\ 
\end{tabular}
\end{ruledtabular}
\caption{The mapping from tensor product state $\ket{a,\mu_a}\otimes \ket{b,\mu_b}$ to a new basis $\ket{c,\mu_c}$ using the $\mathbb{Z}_2$ fusion rule. Degeneracy basis labels $\mu_x$ for each charge $x  \in (a,b,c)$ have been included to count fusion into a particular charge.\label{MappingVectorSpaceZ2}}
\end{table}

However, more structure can be included into the linear map by defining some relationship between the basis labels of the spaces. For example, assume we include a $\mathbb{Z}_2$ fusion rule defined by
\begin{equation}
1 \times 1 \rightarrow 0;   \quad 0 \times a \rightarrow a \quad \forall ~a .
\end{equation}
The charge outcome $c = 1$, resulting from the fusion $0 \times 1$ and $1 \times 0$, is degenerate, as is $c=0$ which results from $0\times 0$ and $1\times 1$. The degenerate outcomes are then indexed by a degeneracy index $\mu_c$. The linear map using the $\mathbb{Z}_2$ fusion rule is given in Table~\ref{MappingVectorSpaceZ2}. 

Therefore, as per \rcite{singh2010}, the linear map tensor can in general be written as
\begin{equation}
T = \bigoplus_{a,b,c} N_{ab}^c\sum_{\mu_a \mu_b \mu_c} 
\left( T_{ab}^c \right)_{\mu_a \mu_b}^{\mu_c} \ket{c,\mu_c} \bra{a,\mu_a} \bra{b,\mu_b} ,
\end{equation}
where the tensor $T$ is constructed blockwise from tensors $T^c_{ab}$, with each block being identified by the charge triple $(a,b,c)$. Each block tensor $T^c_{ab}$ then has its entries indexed by the corresponding degeneracy indices $(\mu_a, \mu_b, \mu_c )$.

\begin{figure}
 \includegraphics{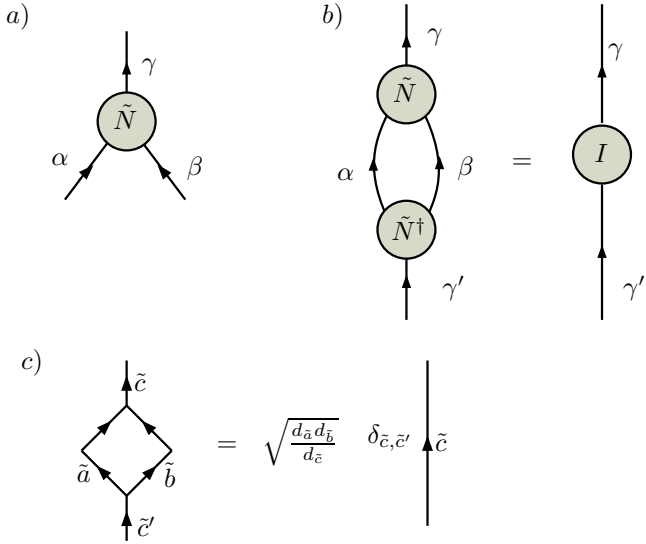}
\caption{a) The diagrammatic representation of the anyonic fusion tensor $\tilde{N}$ which can be expressed in its block structure $\tilde{N}_{\alpha \beta}^{\gamma} = \left( \tilde{N}_{\tilde{a} \tilde{b}}^{\tilde{c}} \right)_{\mu_{\tilde{a}} \mu_{\tilde{b}}}^{\mu_{\tilde{c}}} =\left(\frac{d_{\tilde{c}}}{d_{\tilde{a}} d_{\tilde{b}}} \right)^{1/4}$ for valid map $\alpha \times \beta \rightarrow \gamma$ and zero otherwise .  b) The fusion tensor $\tilde{N}_{\alpha \beta}^{\gamma}$ and its Hermitian conjugate  $\tilde{N}^{\alpha \beta}_{\gamma'}$ are linear and hence their product contracts to the identity operator defined on the new multi-index $\gamma$. c) Elimination of loops from anyonic diagrams as given in Ref.~\onlinecite{Bonderson2007}.}
\label{AnyonicFusionTensor}
\end{figure}

We generalize this to anyonic systems admitting Anyon~$\times$~U(1) symmetries as follows: 
Let two sites of an anyonic system be described by a degenerate Hilbert space $\mathbb{V}^{(A)}$ and $\mathbb{V}^{(B)}$ with basis $\{\alpha = (\tilde{a}, \mu_{\tilde{a}})\}$ and $\{\beta = (\tilde{b}, \mu_{\tilde{b}}) \}$, and let the anyonic fusion product define a ``fusion map'' $\tilde{N}_{\alpha, \beta}^{\gamma}$ from multi-indices $\alpha$ and $\beta$ to a new multi-index $\gamma$. The anyonic fusion map creates a new vertex and we normalize it according to diagrammatic isotopy convention. As was discussed previously, the map is created by iterating slowly over basis label $\alpha$ and fast over $\beta$, and enumerating pairs $(\alpha,\beta)$ by a new label $\gamma$. The fusion tensor is represented in Fig.~\ref{AnyonicFusionTensor}(a). However, unlike the case of abelian symmetry, for anyons normalized according to the diagrammatic isotopic convention the coefficients of a valid fusion map $\alpha \times \beta \rightarrow \gamma$ 
take the value of the vertex normalization factor $\left(\frac{d_{\tilde{c}}}{d_{\tilde{a}} d_{\tilde{b}}} \right)^{1/4}$. As for abelian anyons, the coefficients are zero if there is no valid fusion map.

The anyonic fusion tensor $\tilde{N}_{\alpha \beta}^{\gamma}$ and its Hermitian conjugate, the splitting tensor $\tilde{N}^{\alpha \beta}_{\gamma}$, are linear maps and fulfill the condition that $\tilde{N}^{\alpha \beta}_{\gamma'}  \tilde{N}_{\alpha \beta}^{\gamma} = I_{\gamma'}^{\gamma}$ (Einstein summation convention assumed) which is an identity operator on the new (degenerate) space $\mathbb{V}^{(C)}$, as shown in Fig.~\ref{AnyonicFusionTensor}(b). The loop resulting from the contraction in Fig.~\ref{AnyonicFusionTensor}(b) is eliminated using the relation in Fig.~\ref{AnyonicFusionTensor}(c). It should be noted that the vertex normalization $\sqrt{\frac{d_{\tilde{c}}}{d_{\tilde{a}} d_{\tilde{b}}}}$ in the definition of the fusion tensor $\tilde{N}$ and splitting tensor $\tilde{N}^{\dagger}$ cancels with the loop factors $\sqrt{\frac{d_{\tilde{a}} d_{\tilde{b}}}{d_{\tilde{c}}}}$ and hence the identity matrix operator in Fig.~\ref{AnyonicFusionTensor}(b) does not contain any factor of the quantum 
dimension $d_{\tilde{a}}$ of anyonic charge $\tilde{a}$. Also note that the quantum dimension $d_{\tilde{a}}$ decomposes as the product $d_{\tilde{a}} = d_a d_{n_a}$ where $d_a$ is the anyon quantum dimension and $d_{n_a}$ is the dimension of U(1) charge, which is trivially equal to one.

\subsubsection{Vertical Bends}
Bending a charge line horizontally is trivial, as timelike (i.e.~horizontal) slices of the fusion tree are invariant under topology-preserving deformations. 
However, vertically bending an anyon charge line is non-trivial and involves reversing the orientation of the anyon worldline. The details of how to resolve the vertical bends in terms of F-moves have been given in Ref.~\onlinecite{Bonderson2007} and also in Ref.~\onlinecite{Pfeifer2015}. We do not repeat the derivations here but only mention the minor changes in the presence of U(1) charges.

We adapt the left bend given in Ref.~\onlinecite{Pfeifer2015} to the case of Anyon $\times$ U(1). This is given as
\begin{equation}
 \includegraphics{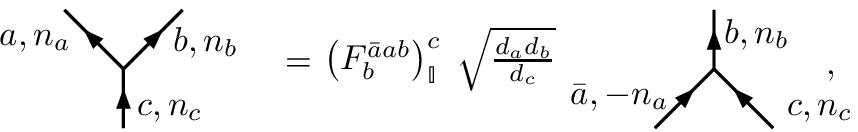}
\end{equation}
where the U(1) charges on the vertex satisfy the condition $n_a + n_b = n_c$. The dual of the anyonic $a$ and that of the U(1) charge $n$ are denoted respectively as $\bar{a}$ and $-n$, which will be the new charge label on the left-bent leg. In addition there is also an implicit F~coefficient from the U(1) charge sector, but this is always equal to~1.
Similarly, the right bend is given by 
\begin{equation}
 \includegraphics[width=\columnwidth]{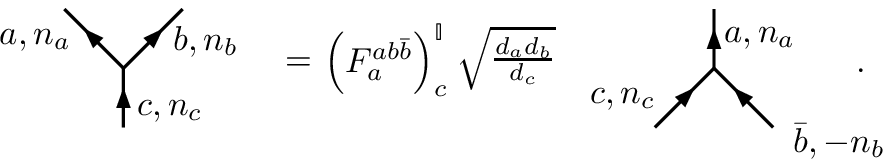}
\end{equation}

In summary, by constructing the appropriate tensor objects with Anyon~$\times$~U(1) symmetry (e.g.~two index and three index tensors, fusion tensors, etc.), one can construct an MPS ansatz for anyonic systems at any fixed rational filling. This ansatz may then be used to construct an an approximation to the ground state of a system by means of anyonic algorithms such as the anyonic TEBD algorithm proposed by Singh et al.\cite{Singh} or the anyonic DMRG.\cite{Pfeifer2015} Ground state properties such as entropy scaling and correlation functions can be computed using approaches similar to those for conventional tensor networks, but modified to account for anyonic statistics by normalizing vertices, removing loops and bending anyonic charge lines in accordance with the prescriptions given in Ref.~\onlinecite{Bonderson2007}.

\section{Test Models and Results}
We provide some examples to demonstrate that an Anyon~$\times$~U(1)-symmetric MPS ansatz may be used to simulate itinerant anyonic systems at any rational filling fraction, and also to provide an example of a tensor network where more than one symmetry is exploited in the algorithm, with one of these symmetries being anyonic. The anyonic models we consider are itinerant hardcore Fibonacci anyons with variable filling fractions and interactions, on a one-dimensional chain (the Golden Chain\cite{Feiguin2007}), and on  a ladder (the Golden Ladder). We compute their ground state energies and entanglement entropies, using the definition of entanglement entropy for non-Abelian anyons given in \rcite{Pfeifer2014a}. From this we extract the central charges of the conformal field theories associated with the infra-red limits of these models. Analytical solutions for these models are not generally known, but we establish the validity of our method by using it to compute equivalent known results for spinless fermions 
and hardcore bosons, and also by comparing results for selected anyonic systems with those obtained 
using anyonic DMRG.\cite{Pfeifer2015} In general, our results are found to be accurate to 4 or 5 decimal places.

\subsection{Itinerant hardcore particles on a one-dimensional chain}
We give some diagnostic test results for hopping and interacting anyons on a chain using an anyonic $t$-$J$ Hamiltonian which is analogous to the $t$-$J$ model for electrons. To make the analogy more apparent, we briefly review the electronic $t$-$J$ model.

\subsubsection{Electronic $t$-$J$ Model}
The electronic $t$-$J$ Hamiltonian consists of two competing terms: a term corresponding to the kinetic energy of the electrons, and an interaction between their spin degrees of freedom. The $t$-$J$ Hamiltonian is
\begin{equation}
\hat{H} = -t \sum_{\langle ij \rangle} \hat{c}_i^{\dagger} \hat{c}_j + J \sum_{\langle ij \rangle} \hat{S}_i \cdot \hat{S}_j ,
\end{equation}
where the first term is the kinetic energy with hopping strength $t$ and $\hat{c}_i^{\dagger}(\hat{c}_i)$ is the creation (annihilation) operator which satisfies fermionic anticommutation relations. The second term is the Heisenberg spin-spin interaction which can be rewritten in terms of projector of nearest spins to the singlet state using the fact that
\begin{equation}
\hat{S}_i \cdot \hat{S}_j  = \frac{1}{2}\left[(\hat{S}_i + \hat{S}_j)^2 - \hat{S}^2_i - \hat{S}^2_j  \right] .
\end{equation}
The addition of two spin-$1/2$ charges is given by the rule,
\begin{equation}
 \frac{1}{2} \otimes \frac{1}{2} = 0 \oplus 1.
\end{equation}
Let $\hat{S} = \hat{S}_i + \hat{S}_j$ and choose units such that $\hbar=1$. Then the relation
\begin{equation}
 \hat{S}^2 \ket{s, m} = s(s+1) \ket{s, m},
\end{equation}
means $\hat{S}^2$ has two eigenvalues, $0$ (when $s=0$) and $2$ (when $s=1$). Therefore $\hat{S}^2$ can be written in terms of projectors to the singlet and triplet subspaces as $(\hat{S}_i + \hat{S}_j)^2 = 0 \hat{\pi}^{(0)}_{ij} + 2 \hat{\pi}^{(1)}_{ij}$, where $\hat{\pi}^{(0)}$ and $\hat{\pi}^{(1)}$ are the projectors to singlet and triplet subspaces. Therefore,
\begin{equation}
\hat{S}_i \cdot \hat{S}_j = -\hat{\pi}^{(0)}_{ij} + \frac{1}{4} ,
\end{equation}
where the identity, $\mathbb{I} = \hat{\pi}^{(0)} + \hat{\pi}^{(1)}$ has been used in the last step. Therefore, the $t$-$J$ Hamiltonian simplifies to 
\begin{equation}
\hat{H} = -t \sum_{\langle ij \rangle} \hat{c}_i^{\dagger} \hat{c}_j - J\sum_{\langle ij \rangle} \pi_{ij}^0 + \text{const.}
\end{equation}
For $J>0$ the Hamiltonian favours neigbouring spins forming singlets (antiferromagnetic), and for $J<0$ it favours triplet formation (ferromagnetic). We adapt the electronic $t$-$J$ to anyons.

\subsubsection{Anyonic $t$-$J$ Hamiltonian in 1D: Hopping term}
Anyonic operators are written as matrices on the fusion space of the participating anyons. A local two-site Hamiltonian $\hat{H} = \sum_i H^{[i,i+1]}$ can be written diagrammatically as
\begin{equation}
 \includegraphics{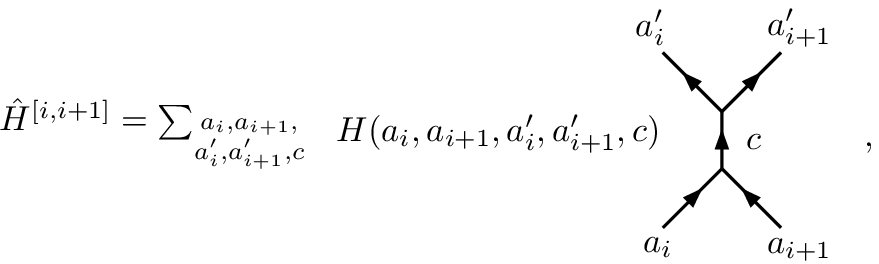}
\end{equation}
where the values of the function $H(a_i, a_{i+1}, a'_i, a'_{i+1}, c)$ are determined by the model being constructed. The conservation of charge $c$ resulting from fusion puts the Hamiltonian into block-diagonal form as $\hat{H}^{[i,i+1]} = \bigoplus_c \hat{H}_c^{[i,i+1]}$. 

The anyonic $t$-$J$ Hamitonian consists of two terms: a hopping term and an interaction term. To give a systematic and concrete treatment of both terms, We give the explicit construction for Fibonacci anyons. 

The hopping of a Fibonacci anyon in 1D means the neigbouring site has to be vacant, corresponding to the vacuum charge $\mathbb{I}$. The kinetic operator can thus be represented as
\begin{equation}
 \includegraphics{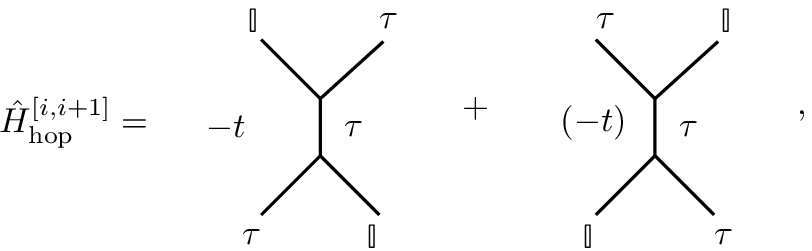}
\end{equation}
being analogous to the fermionic terms $( \hat{c}_{i \sigma}^{\dagger} \hat{c}_{i+1, \sigma} + \text{h.c} )$ that translate a fermion between sites $i$ and $i+1$. Since the anyonic hopping term requires there is a vacant site with vacuum charge $\mathbb{I}$, this implies that the hopping term is nonzero only when $(a_i=\mathbb{I}, a_{i+1}=\tau, a'_i=\tau, a'_{i+1}=\mathbb{I})$ or when $(a_i=\tau, a_{i+1} = \mathbb{I}, a'_i=\mathbb{I},a'_{i+1}=\tau)$. For dynamics in one dimension with hardcore constraints, the underlying exchange statistics of the particle do not affect the ground  state properties (though the degeneracy of the ground states may differ for different particle species). Therefore, itinerant hardcore Fibonacci anyons, spinless fermions and hardcore bosons all have the same ground state energies at any rational filling of the lattice. 

\begin{figure}
 \includegraphics[scale=0.7]{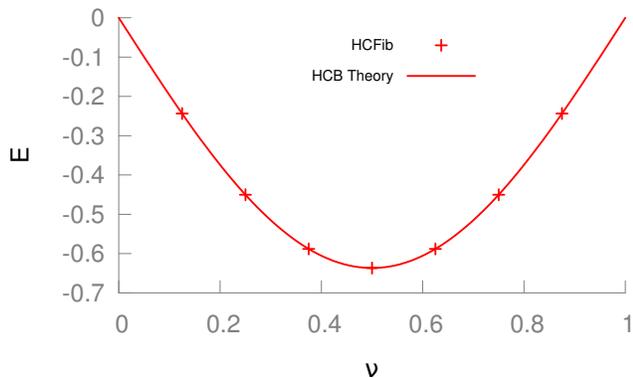}
\caption{(Color online) Ground state energy ($E$) of itinerant hard-core Fibonacci anyons on an infinite chain at different filling fractions ($\nu$). The data points result from numerical simulations, while the smooth curve is the ground state energy for an equivalent system of spinless fermions at the thermodynamic limit.}
\label{HCBChainEnergy}
\end{figure}

We use the Anyon $\times$ U(1) symmetric TEBD algorithm to compute the ground state energies of itinerant Fibonacci anyons, spinless fermions and hardcore bosons on a 1D lattice. We obtained the same ground state energies for these three cases up to $4$ to $5$ decimal places. This is owing to the fact that particles are not allowed to exchange positions on the lattice, and thus particle statistics do not affect the ground state propoeties. 

In Fig.~\ref{HCBChainEnergy} we plot the numerical ground state energy of itinerant Fibonacci anyons against the analytical ground state energy of an equivalent system of spinless fermions given by
\begin{equation}
E(t, \nu) = -2t \frac{\sin(\pi \nu)}{\pi} .
\end{equation}
Numerical ground state energies for spinless fermions and hardcore bosons result in an identical curve and so are not reproduced here. 

\subsubsection{Anyonic $t$-$J$ Hamiltonian in 1D : Heisenberg interaction term}
Next we include an anyonic Heisenberg interaction term in addition to the hopping term. The anyonic Heisenberg interaction is constructed by analogy to the Heisenberg spin-spin interaction. For 100\% filling this model was first proposed and studied by Feiguin et~al.,\cite{Feiguin2007} and is known as the Golden Chain. The anyonic Heisenberg interaction takes the form
\begin{equation}
\includegraphics{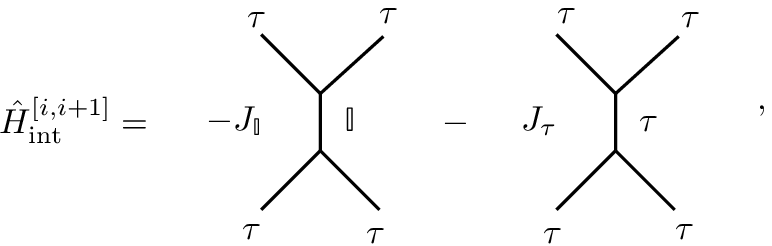} 
\end{equation}
where $J_{\mathbb{I}} > J_{\tau} $ corresponds to an antiferromagnetic interaction favouring fusion of the two Fibonacci anyons to the vacuum charge $\mathbb{I}$, and $J_{\mathbb{I}}<J_{\tau}$ corresponds to a ferromagnetic interaction favouring projection to the Fibonacci charge $\tau$. 

When a Heisenberg interaction is introduced into a system of itinerant Fibonacci anyons, the extensive degeneracy of the free anyon system is lifted. The Hilbert space of the interacting itinerant anyon system admits the decomposition
\begin{equation}
\mc{H} = \mc{H}_\mrm{config}\otimes\mc{H}_\mrm{fusion}
\end{equation}
where $\mc{H}_\mrm{config}$ is the space of particle configurations, and $\mc{H}_\mrm{fusion}$ is the space of valid labelings of the fusion tree. The Hamiltonian admits an equivalent decomposition, and the Hamiltonian for a system of free particles (acting on $\mc{H}_\mrm{config}$ is associated with 
a central charge of 1. When a Heisenberg-type interaction is added, this acts on $\mc{H}_\mrm{fusion}$, lifting the degeneracy of the states in this subspace. For a critical interaction, the total central charge is additive, and may be written $1+c$ where 1 is the contribution from the itinerant anyon model acting on $\mc{H}_\mrm{config}$ and $c$ is the contribution from the interactions on the fusion portion of the Hilbert space.\cite{Poilblanc2011a}
This is alluded to as spin charge separation. From our numerical simulations, when $J_\mathbb{I}>J_{\tau}$ (antiferromagnetic), we obtained $c = 0.708$ and when $J_\mathbb{I}<J_{\tau}$ (ferromagnetic), we obtained $c=0.84$, for total central charges of $1.708$ and $1.84$. These are very close to the expected central charges of $1+7/10$ for antiferromagnetic interaction and $1+4/5$ for ferromagnetic interaction.

\subsection{Anyonic $t$-$J$ Model on ladder}
Nonabelian anyons have nontrivial braid factors, making their simulation difficult. 
For such systems, numerical approaches based on Monte Carlo schemes are plagued by a form of the sign problem. We offer numerical evidence that anyonic tensor networks (such as the anyonic MPS) are able to simulate anyonic systems on geometries beyond one dimension, in situations where the anyons experience braiding. To model how braiding statistics affect the ground state of anyons, we introduce the anyonic $t$-$J$ model on the ladder, as a generalisation of the model already considered on a chain. Each site on the ladder supports only two types of charges, namely, either a vacuum charge $\mathbb{I}$ or a single Fibonacci anyon $\tau$. Unlike in one dimension, anyons on the ladder can exchange positions and consequently braid. 

\begin{figure}
 \includegraphics[width=\columnwidth]{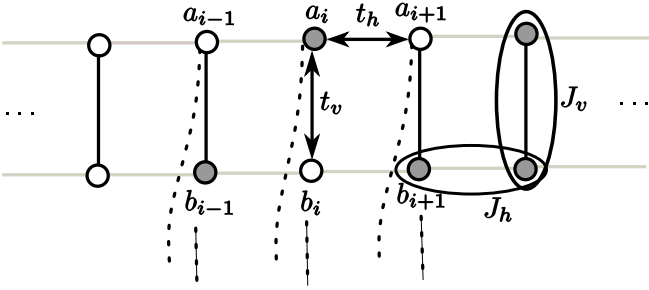}
\caption{A ladder of itinerant anyons. Ficticious strings are attached to each nontrivial charge to indicate that they can participate in nontrivial braids as they exchange positions with neigbouring anyons. For example, the nontrivial anyonic charge $a_i$ braids with the nontrivial anyonic charge $b_{i-1}$ as it hops horizontally to a new site with $a_{i-1}=0$. The labels $J_v$ and $J_h$ represent the amplitudes for projecting the corresponding circled pairs into the vacuum sector.} 
\label{AnyonicLadder}
\end{figure}

To model the braiding of anyons on the ladder in a consistent manner, we impose a linear ordering to the anyons by attaching ficticious ``strings'' to the anyons and oriented them leftward of their on-site position (see Fig.~\ref{AnyonicLadder}). When an anyon hops from one site to another on either the top or bottom chain of the ladder, it braids with any adjacent anyonic charge along its trajectory, with the strings acting as a convenient mnemonic to visualise the orientation of the braid.

With reference to Fig.~\ref{AnyonicLadder}, the anyonic $t$-$J$ Hamiltonian can be written as
\begin{align}
\hat{H} & = -t_h \sum_{i=1}^{N-1} \left( \hat{b}_{a_i \rightarrow a_{i+1}=\mbb{I}}  + \hat{b}_{b_i \rightarrow b_{i+1}=\mbb{I}} + \text{h$\cdot$c} \right) \nonumber \\
& \quad \quad \quad \quad - J_h \left( \hat{\Pi}^{\mathbb{I}}_{a_i, a_{i+1}} + \hat{\Pi}^{\mathbb{I}}_{b_i, b_{i+1}} \right) \nonumber \\
& \quad \quad - \frac{t_v}{2} \sum_{i=1}^N \left( \hat{b}_{a_i \rightarrow b_i=\mbb{I}} + \hat{b}_{a_{i+1} \rightarrow b_{i+1}=\mbb{I}} + \text{h$\cdot$c} \right) \nonumber \\
& \quad \quad \quad \quad - \frac{J_v}{2} \left( \hat{\Pi}^{\mathbb{I}}_{a_i, b_i} + \hat{\Pi}^{\mathbb{I}}_{a_{i+1}, b_{i+1}} \right) ,
\end{align}
where $(t_h, t_v)$ and $(J_h, J_v)$ are the hopping and interaction amplitudes for anyons on the legs and rungs of the ladder. The vacuum charge is denoted by $\mbb{I}$. The operator $\hat{b}_{x \rightarrow y=\mbb{I}}$ moves a nontrivial charge $x$ into a new site having trivial vacuum charge $y=\mbb{I}$ while it braids the charge $x$ with any other charge along its path. The projector $\hat{\Pi}^\mathbb{I}_{x,y}$ projects the nontrivial anyonic charges $x$ and $y$ into a vacuum charge $\mbb{I}$. The anyonic interaction is antiferromagnetic when $J>0$ and the interaction becomes ferromagnetic when $J<0$. The Hamiltonian along the rung has been symmetrized with half a contribution from each of the rungs on sites $i$ and $i+1$. 

\begin{figure}
 \includegraphics{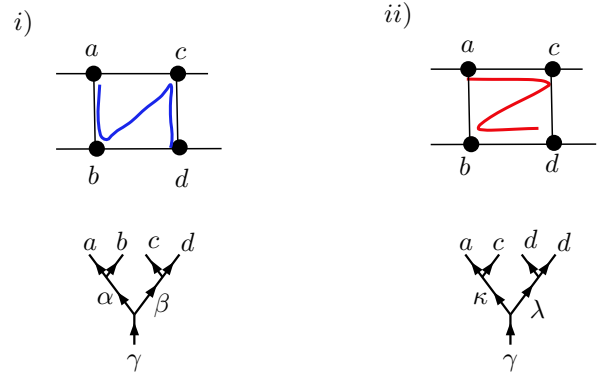}
\caption{(Color online) The two convenient fusion orderings, with their respective fusion trees shown underneath. The first fusion order	couples charges $(a,b)$ and $(c,d)$ while the second couples charges $(a,c)$ and $(b,d)$.}
\label{FusionOrderPlaquette}
\end{figure}

Below, we show an explicit derivation of the Hamiltonian terms which can be arranged as a charge-conserving matrix operator. The local Hamiltonian $\hat{h}$ is derived on a plaquette whose vertices are labeled $(a,b,c,d)$ for brevity as shown in Fig.~\ref{FusionOrderPlaquette}. The local Hamiltonian is written as 
\begin{align}
 \hat{h} &= -t_h \left( \hat{b}_{a \rightarrow c=\mbb{I}}  + \hat{b}_{b \rightarrow d=\mbb{I}} + \text{h$\cdot$c} \right)  - J_h \left( \hat{\Pi}^{\mathbb{I}}_{a,c} + \hat{\Pi}^{\mathbb{I}}_{b,d} \right) \nonumber \\
& \quad - \frac{t_v}{2} \left( \hat{b}_{a \rightarrow b=\mbb{I}} + \hat{b}_{c \rightarrow d=\mbb{I}} + \text{h$\cdot$c} \right) -  \frac{J_v}{2} \left( \hat{\Pi}^{\mathbb{I}}_{a,b} + \hat{\Pi}^{\mathbb{I}}_{c,d} \right) .
\end{align}

Depending on the imposed fusion order, some of the operators will be diagonal in the fusion basis. The two most convenient fusion order are shown in Fig.~\ref{FusionOrderPlaquette}. Let the first basis be denoted as $\ket{\text{I}} = \ket{(ab;\alpha)(cd;\beta)(\alpha \beta; \gamma)}$ with the fusion order $((a,b)(c,d))$ where the anyons $(a,b)$ and $(c,d)$ are first fused independently, then fuse their outcomes and let the second basis be $\ket{\text{II}} = \ket{(ac;\kappa)(bd;\lambda)(\kappa \lambda; \gamma)}$ with fusion order $((a,c)(b,d))$. Using a series of $F$-moves and $R$-moves, the first basis transforms into the second basis according to
\begin{equation}
 \includegraphics{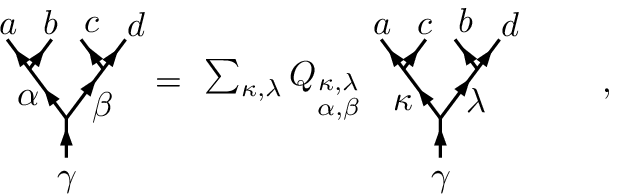}
\end{equation}
where the tensor $Q_{\alpha, \beta}^{\kappa, \lambda}$ is given by,
\begin{equation}
Q_{\alpha, \beta}^{\kappa, \lambda} = \sum_{\eta, \theta} 
\left[\left(F^{\alpha c d}_{\gamma} \right)^{-1}\right]^{\eta}_{\beta} 
\left(F^{a b c}_{\gamma} \right)^{\theta}_{\alpha} R^{bc}_{\theta}
\left[\left(F^{a c b}_{\gamma} \right)^{-1}\right]^{\kappa}_{\theta} 
\left(F^{\kappa b d}_{\gamma} \right)^{\lambda}_{\eta} ~
\end{equation}
with its derivation being given in Appendix~\ref{FusionBasisTransform}.

The rest of the derivation of the matrix expression for the Hamiltonian $\hat{h}$ is performed as an illustrative example in Appendix~\ref{HamiltonianAnyonicLadder}.

\subsubsection{Itinerant Fibonacci anyons, spinless fermions and hardcore bosons on a ladder}
We exploit the anyonic and U(1) symmetries of the model both in the MPS ansatz and in the Hamiltonian $\hat{H}$, and use the TEBD algorithm to compute the ground state energies of itinerant Fibonacci anyons on the ladder at different filling fractions. Since the MPS has a one-dimensional structure, we map the ladder to a chain by fusing the anyonic charges on each rung to make a new single site. The vertical and the horizontal hopping rates are set equal to one, $t_h = 1$ and $t_v=1$ while the vertical and horizontal Heisenberg interactions $J_v$ and $J_h$ are set to zero. 

\begin{figure}
\includegraphics[scale=0.7]{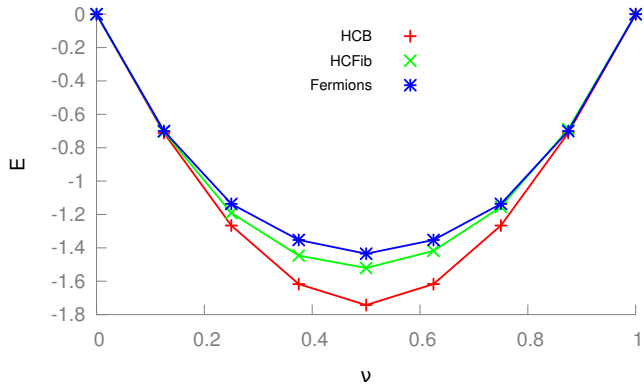}
\caption{(Color online) The ground state energies of hardcore bosons (HCB), spinless fermions and itinerant Fibonacci anyons (HCFib) on a two-leg ladder when only hopping is turned on. The line is a guide to the eye.}
\label{ItinerantModels}
\end{figure}

There are no known analytical results for the ground state of itinerant Fibonacci anyons on a ladder, but we test the validity of our method against the ground state energies of itinerant hardcore bosons and spinless fermions on the ladder shown in Fig.~\ref{ItinerantModels}. The phase diagram of this model for unit filling fraction was studied in Ref. \onlinecite{PhysRevB.83.134439}.

It can be seen from the figure that incorporating the capacity for anyons to braid around one another results in an increase in the ground state energy per particle. This fact is reminiscent of the property that a system of identical fermions have a higher energy than bosons due to Pauli exclusion principle in real space. This also implies that there might exist a Pauli-like exclusion principle for anyons too, at least in some regimes.\cite{Pfeifer2015a} We also see from the figure that while the bosons and fermions have a paricle-hole symmetry which is reflected in the symmetric ground state energy around half-filling $\nu = \frac{1}{2}$, the system of Fibonacci anyons on the ladder does not display this symmetry. {One of the consequences of particle-hole symmetry is that the ground state energ{ies} $E$ at filling fraction{s} $\nu$ and $1-\nu$ should be equal. While this is known for fermions and bosons, and {reproduced by} our numerical results 
as shown in Table.~\ref{GroundStateEnergyItinerantModels}, {we see from our numerical results} that this no longer holds for {some} non-abelian anyon model such as Fibonacci anyons, though {in this instance} the breakdown of particle-hole duality is weak {in the sense that it has only a very small impact on ground state energies}. {Interference of braiding particles} raises the ground state energies, and thus {the} higher filling fraction{s} $\nu{>1/2}$, e.g. $\nu=5/8$, {have} slightly higher energ{ies} than the $1-\nu$ state{s}, e.g. $\nu=3/8$.}

\begin{table}
\begin{ruledtabular}
\begin{tabular}{  c  c  c  c }
$\nu$   &  $E_{\text{HCB}}$  &  $E_{\text{HCFib}}$      &  $E_{\text{SF}}$    \\ \hline
0     &        0                &          0            &         0           \\ 
1/8   &    -0.71162             &      -0.70397         &     -0.70015        \\  
2/8   &    -1.26597             &      -1.19102         &     -1.13658        \\ 
3/8   &    -1.61707             &      -1.44620         &     -1.35273        \\ 
4/8   &    -1.74300             &      -1.52085         &     -1.43534        \\
5/8   &    -1.61707             &      -1.41803         &     -1.35271        \\
6/8   &    -1.26597             &      -1.15486         &     -1.13660        \\
7/8   &    -0.71162             &      -0.68857         &     -0.70015        \\
1     &       0                 &          0            &         0           \\
\end{tabular}
\end{ruledtabular}
\caption{The values of the ground state energy $E$ at various filling fractions $\nu$ corresponding to figure Fig.~\ref{ItinerantModels}. The subscripts in $E_{(\bullet)}$ are ``HCB'' for hardcore bosons, ``HCFib'' for hardcore Fibonacci anyons, and ``SF'' for spinless fermions. The values are given to five decimal places. The ground state energies of bosons and fermions are symmetric around half-filling, but not so for Fibonacci anyons.}
\label{GroundStateEnergyItinerantModels}
\end{table}

{The origin of the breakdown in the particle-hole duality is in the difference {in} the fusion degrees of freedom of the particle types. For systems of bosons or fermions, the fusion space is one-dimensional, independent of the number of particles. In contrast, for non-Abelian models such as the Fibonacci model, the fusion space grows exponentially with 
the number of anyons and hence is not symmetric under particle hole exchange. Braiding acts non-trivially on the fusion degrees of freedom and changes the ground state energy in way that is not particle hole symmetric.}

\subsubsection{Phase diagram of the Golden Ladder}
\begin{figure}
\includegraphics[scale=0.8]{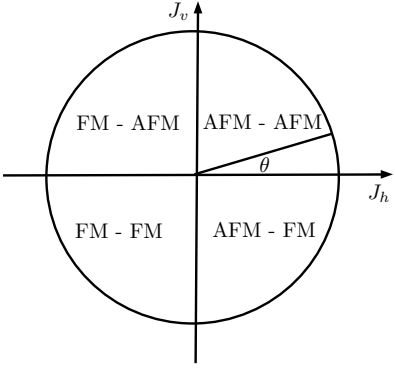}
\caption{The horizontal and vertical interaction stengths $(J_h, J_v)$ on the legs and rungs of the ladder are parameterized by $(\cos{\theta}, \sin{\theta})$ along the legs and rungs respectively. The labels within each quadrant indicate the nature of the interaction, whether antiferromagnetic or ferromagnetic.} 
\label{InteractionStrengthParamaterization}
\end{figure}

{We further test our ansatz by studying the entanglement structure{s of} ground states of interacting Fibonacci anyons on the ladder at unit filling. This model has been studied in Ref.~\onlinecite{PhysRevB.83.134439}, and we {verify our ansatz by reproducing known phases of the model at specific values of the tunable parameters}}. At unit filling, there is a single localized  Fibonacci anyons per site of the ladder and therefore hopping rates are everywhere zero. This is a quasi-1D generalisation of the Golden Chain,\cite{Feiguin2007} which might be called the Golden {L}adder. The relative interaction strengths of the legs and rungs of the ladder, including both antiferromagnetic and ferromagnetic couplings, may be parameterized on a circle (see Fig.~\ref{InteractionStrengthParamaterization}, where the {ferromagnetic or antiferromagnetic natures of the interactions} in each sector are indicated). 

{We evolve this model to ground state using TEBD, and compute the scaling of the block entanglement entropy from von~Neumann's relation, 
\begin{equation}
S(r) = -\text{Tr}(\hat{\rho}_r ~\text{log} ~\hat{\rho}_r) ,
\end{equation}
where $\hat{\rho}_r$ is in general the reduced density matrix of a block of $r$ sites, here $r$ rungs. {From conformal field theory, the scaling of entanglement entropy on a system with an open boundary is}
\begin{equation}
 S(r) = \frac{c}{3} \log{r} ,
\end{equation}
where $c$ is the central charge of the system at criticality. This relation means that, for a critical model, the entanglement block scaling---computed from the MPS ground state representation---should display a logarithmic relation with the block size. The central charge $c$ can {then} be extracted from the {relationship}}
\begin{equation}
{ c = 3\, \frac{S(r_2) - S(r_1)}{\log{r_2} - \log{r_1}}}.\label{eq:cch}
\end{equation}

\begin{figure}
\includegraphics[width=\columnwidth]{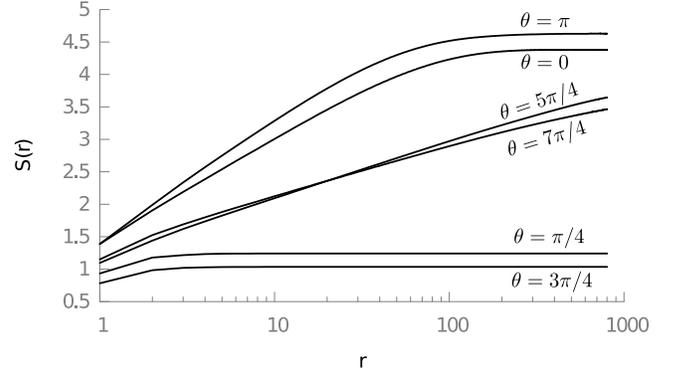}
\caption{Scaling of entanglement entropy $S$ as a function of block size $r$, for different angles on the circle ($\theta$), which correspond to different ratios of coupling strength between the legs and rungs of the ladder.}
\label{InteractingFibonacciLadder}
\end{figure}

\begin{figure}
\includegraphics[scale=0.8]{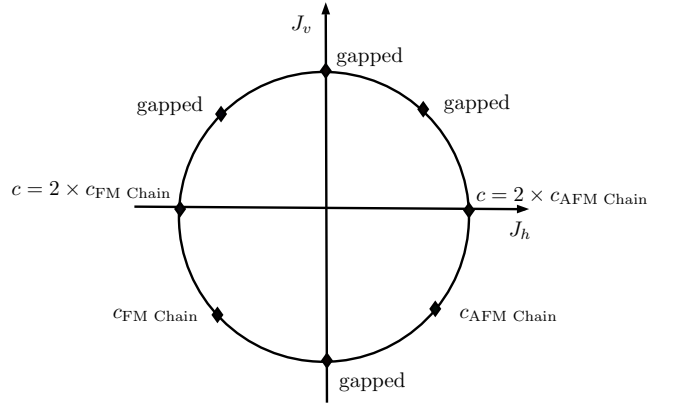}
\caption{The central charge of the underlying CFT extracted from the scaling of the entanglement entropy of Fig.~\ref{InteractingFibonacciLadder} are shown at the paremeter points we considered. When vertical coupling is set to zero and $J_h$ range from $-1$ to $+1$, we obtain central charge which doubles that of single critical FM or AFM chains which lies on the equator. The values of $c_{\text{AFM Chain}}$ and $c_{\text{FM Chain}}$ are given as $c_{\text{AFM Chain}} = 0.704$ and $c_{\text{FM Chain}} = 0.801$. {Phase boundaries for this model may be found in} {Fig.~11(a) of Ref.~\onlinecite{PhysRevB.83.134439}.}
}
\label{CentralChargePieChart}
\end{figure}

The block entanglement entropy for various parameter regimes are shown in Fig.~\ref{InteractingFibonacciLadder}, and their central charges are indicated in Fig.~\ref{CentralChargePieChart}.
{As seen in Fig.~\ref{InteractingFibonacciLadder} the finite bond dimension of the MPS causes entanglement to artificially plateau over larger distances $r=|r_2-r_1|$, but calculation of $c$ using Eq.~(\ref{eq:cch}) may be performed for any separation $r$ prior to this plateau, where an appropriate linear correlation is obtained between $S(r_2)-S(r_1)$ and $\log{r_2}-\log{r_1}$.}

 One can interpret this Fig.~\ref{CentralChargePieChart} by considering how the physics of the interacting Fibonacci anyon changes as the parameterization angle $\theta$ is varied. When $\theta=0$, there are no couplings along the rungs and we have $2$ chains of Fibonacci anyons with antiferromagnetic interactions. The system in this parameter regime is gapless and has a central charge which is twice that of a single chain, i.e. $2 \times 7/10$. Even though the MPS most naturally yields exponentially decaying correlators, we are nevertheless able to extract an approximate value for the central charge, $c = 1.405$, from the linear part of the curve. When $\theta = \pi/4$ and $\theta = 3\pi/4$, the vertical couplings are antiferromagnetic favouring pairs of Fibonacci anyon fusing into the vacuum charge. This phase is gapped with central charge $c=0$. 
When $\theta=\pi/2$, the Hamiltonian favours fusion of pair of $\tau$ charges on each rung to the vacuum charge, and is hence a product state which is unique and gapped. The phase is not critical  and has a central charge $c=0$. At the $\theta=\pi$ point, the horizontal coupling $J_h=-1$ is ferromagnetic, while the vertical coupling $J_v$ is zero, and the ladder reduces to two copies of a ferromagnetic Golden Chain. From our numerical simulation, we computed a central charge of $c=1.629$ which is close to the expected theoretical value of $c=2\times 4/5$. At the point $\theta = 5\pi/4$, the horizontal and vertical couplings are ferromagnetic. Fusion of the $\tau$ charges on the rungs and legs favours projection to the $\tau$ channel (triplet state). This can easily be pictured by considering a linearized version of the ladder. Nearest neighbour $\tau$ charges on the rung becomes nearest neighbour on the chain and nearest neighbours on the legs becomes next-to-nearest neighbour on the chain.\cite{Trebst2008a} 
Heuristically, fusion to the $\tau$ fusion channel makes the ladder effectively like a single Fibonacci chain and therefore has the same central charge as a single chain. We obtain a central charge of $c=0.801$ which is close to the expected $c=4/5$. When $\theta=\frac{3\pi}{2}$, the vertical coupling $J_v = -1$ is ferromagnetic while the horizontal coupling is zero. This favours projection of neighbouring $\tau$ charges on the rungs into the $\tau$ channel. The ladder reduces to a chain of decoupled $\tau$ charges which has an exponentially large degeneracy in intermediate fusion degrees of freedom. Hence a generic ground state at this point obeys a volume law rather than area law. This system is gapped and not described by conformal field theory. The parameter point $\theta=7\pi/4$ correspond to horizontal antiferromagnetic coupling on the leg and vertical ferromagnetic coupling which is effectively an antiferromagnetic interacting chain. The obtained central charge is $c=0.704$, being close to the 
expected value of $c=7/10$. {Our findings are in agreement with known results showing that the entire upper semicircle is gapped while the lower semicircle is gapless} with the exception of the indicated point at $\theta=3\pi/2$. 

{{Table~\ref{CentralChargesTable} compares} the extracted central charges with their expected theoretical values.}

\begin{table}
\begin{ruledtabular}
\begin{tabular}{  c  c  c }
$\theta$        &       $c_{\text{sim}}$    &        $c_{\text{Theo.}}$               \\ \hline
$0$             &            $1.405$        &          $2 \times 7/10$                 \\ 
$\pi/4$         &             $ 0 $         &             0                            \\  
$\pi/2$         &             $ 0 $         &             0                             \\ 
$3\pi/4$        &             $ 0 $         &             0                             \\ 
$\pi$           &             $1.629$       &         $2\times 4/5$                     \\
$5\pi/4$        &             $0.801$       &            $4/5$                          \\
$3\pi/2$        &              $0$          &             $0$                           \\
$7\pi/4$        &             $0.704$       &             $7/10$                                              \\
\end{tabular}
\end{ruledtabular}
\caption{The table shows the obtained central charges $c_{\text{Sim.}}$ from our numerical simulations, compared against their theoretical values $c_{\text{Theo.}}$ known from conformal field theory, at the interaction strengths parameterized by $\theta$ according to Fig.~\ref{InteractionStrengthParamaterization}. The values correspond to the points shown in Fig.~\ref{CentralChargePieChart}. Where a model is not critical, and hence not described by CFT, we have substituted their central charge $c$ with zero. 
\label{CentralChargesTable}}
\end{table}

\section{Conclusion}
In this paper we show how the anyonic tensor network formalism of Refs.~\onlinecite{Pfeifer2010a, Singh, Pfeifer2015} may be applied in the context of particles admitting multiple charge labels, specifically an anyonic charge and a U(1) charge, here corresponding to particle number. We constructed test models involving both hopping and interaction terms, with this construction being explicitly elaborated in the Appendix. Application of the anyonic infinite TEBD algorithm\cite{Singh} permitted calculation of the ground states of these systems, their entanglement entropies, and central charges. In doing so, we successfully reproduced elements of the phase diagrams for these systems which have previously been obtained using exact diagonalisation.\cite{Feiguin2007,Trebst2008, Trebst2008a, Poilblanc2011a,Poilblanc2011}

This paper consequently demonstrates the feasibility of applying anyonic TEBD to systems of particles admitting both anyonic and U(1) conserved charges. The method presented here can be used to probe new regimes of the physics of anyons such as equilibrium phases of quasi-1D systems of braiding anyons at arbitrary density as well as non-equilibrium dynamics of anyons in two dimensional systems at low density. The later could be used to study the robustness of large size topological quantum computers/memories to errors induced by \emph{coherent} propagation of erroneous anyons created by thermal fluctuations which braid around logical degrees of freedom.


\acknowledgments
This research was supported in part by the ARC Centre of Excellence in Engineered Quantum Systems (EQuS), Project No. CE110001013.

\bibliography{MPSBraidedAnyons}

\appendix 

\section{Basics of Anyonic Tensor Networks}\label{AnyonicTensorNetworks}
Using tensor networks to simulate quantum systems involves choosing a network pattern of connected tensors along with a choice of an algorithm that optimizes the representation of the many body state.\cite{Orus2013} In what follows, we review the basic objects common to most anyonic tensor networks (TN). In the main text, we used these objects to construct the anyonic MPS, and our modified Anyonic-U(1) MPS ansatz. 

\subsection{Components of anyonic tensor networks}
In the discussion below, we assume some familiarity with theory of anyons as described in Refs.~\onlinecite{Kitaev2008, Bonderson2007,bonderson2008}. For a pedagogical introduction to anyons, see e.g.~Refs. ~\onlinecite{Preskill1999,Pachos2012}.

The basic objects in any tensor network include vectors (or one-index tensors),  matrices (or two-index tensors), and more generally, $n$-index tensors. We start by examining how the basis of states for a system of anyons can be enumerated, and how it is used in constructing the anyonic equivalents of the above-named TN objects. 
                                                                                                                 
\subsubsection{Anyonic basis enumeration and fusion lookup tables}
The basis of the Hilbert space of anyons is described by a labeled directed fusion tree (see Fig.~\ref{FusionTree}) where the charge $c$ on any incoming edge at a vertex is determined from the charges $a$ and $b$ of the two outgoing edges around the same vertex, according to the fusion rules of the anyon theory
\begin{equation}
 a \times b \rightarrow \sum_c N_{ab}^c ~ c ,
\end{equation}
which implies that charges $a$ and $b$ are allowed to fuse to possibly several different charges $c$. For example, in the fusion tree in Fig.~\ref{FusionTree}(ii) charges $b_1$ and $a_3$ fuse into all possible charges $b_2$. We restrict to multiplicity-free anyon models in this paper, i.e.~$N_{ab}^c \in\{0,1\}$, which includes some of the models most relevant to current experiment such as Ising anyons and Fibonacci anyons.
The iterative fusion process described by a fusion tree makes the mathematical description of the collective state space to be non-local and hence not naturally reducible to a tensor product space of individual anyonic degrees of freedom. 

The description of labeled fusion trees can become extremely verbose in the limit of large system sizes. The amount of data needed to specify the labeling of a fusion tree can be greatly reduced if one only enumerates the labelings having a particular total charge at the trunk. To this end, let $c$ be total charge at the trunk of the fusion tree [see Fig.~\ref{FusionTree}(i)] and assign an index $\mu_c = 1, 2, \cdots, \nu_c$ in increasing numerical order to each unique labeling of the fusion tree.  The index $\mu_c$ is called the degeneracy index and $\nu_c$ is called the degeneracy of charge $c$.\cite{Pfeifer2010a} All the fusion trees are therefore concisely labeled by the multi-index $\gamma=(c, \mu_c)$, with $c$ as the total charge label and $\mu_c$ as its degeneracy index.

This assignment can be described using a ``fusion lookup table'' where the sets of charges corresponding to unique labelings of the fusion tree are recorded in rows, and each row is additionally labeled with (i)~the total charge $c$ and (ii)~an index value $\mu_c$ such that each pair $(c,\mu_c)$ is unique. The use of fusion lookup tables serve a two-fold purpose in anyonic tensor networks. First they can serve as a ``cache", allowing for a total recovery of all the charges labeling a particular fusion tree which may be useful when performing diagrammatic (topological) manipulations of the fusion tree. Secondly, they can be used to significantly reduce the computational cost of contracting an anyonic TN, a fact which has previously been discussed in the context of non-anyonic symmetric TNs (see Refs.~\onlinecite{Singh2011, Singh, Pfeifer2015} for more details). When constructing a fusion table, it is required that there be a one-to-one correspondence between the charge labelings of the fusion tree and the 
charge/degeneracy multi-indices which are assigned to these labelings.

\subsubsection{Anyonic state vector}
\begin{figure}
 \includegraphics[width=\columnwidth]{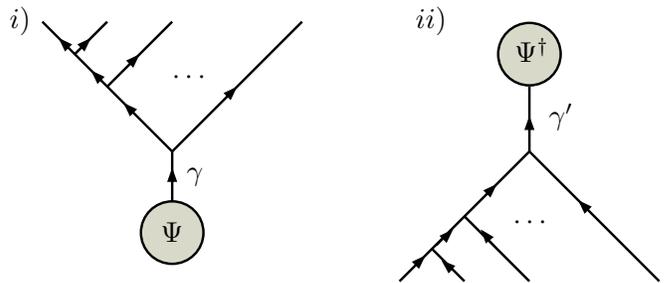}
\caption{(i)~Anyonic state vector and (ii)~its Hermitian conjugate.}
\label{AnyonicQuantumState}
\end{figure}

An anyonic quantum state $\ket{\Psi}$ can be written as a weighted superposition of all labelings of a fusion tree having a total vacuum charge $\mathbb{I}$.\footnote{A system of anyons which does not fuse to the total vacuum charge $\mathbb{I}$ can be considered to be within a larger system with a nontrivial boundary charge that enforces a complete annihilation of the anyonic charges to the vacuum charge.} More compactly, in the multi-index notation, the quantum state can be written as
\begin{equation}
\ket{\Psi} = \sum_{\gamma} \Psi^{\gamma} \ket{\gamma} , 
\end{equation}
where $\gamma =(\mathbb{I},\mu_{\mathbb{I}})$ is an index enumerating all the valid fusion trees. If all the enumerated fusion trees are associated with normalized anyonic diagrams, which is referred to as the \emph{implicit normalization} scheme, then the state amplitudes $\Psi^{\gamma}$ can be arranged as a column vector in the standard basis. Following the diagrammatic notations employed for anyonic tensors in Ref.~\onlinecite{Pfeifer2010a}, we depict the anyonic quantum state by a filled circle with a central leg enumerating all the multi-indexed bases, and an unlabeled tree structure, as shown in Fig.~\ref{AnyonicQuantumState}(i). 
It should be noted that if topological manipulations were to be performed on the fusion tree, such as vertically bending a line opposite to its orientation, it is preferable to use diagrammatic isotopy convention. In such a case, we adopt the prescription given in Ref.~\onlinecite{Bonderson2007}, where the fusion diagrams are weighted with certain pre-factors of quantum dimensions of the anyonic charges on the fusion tree. For the fusion basis shown in Fig.~\ref{FusionTree}, the pre-factor would be $\left(\frac{d_c}{d_{a_1}d_{a_2}\cdots d_{a_k}}\right)^{1/4}$. These normalization factors are then also absorbed into the amplitudes defining the anyonic state vector. This is referred to as the \emph{explicit normalization} scheme. During topological manipulations, all the charges labeling a particular fusion tree can be recovered from the fusion lookup tables and used in computing the necessary data associated to that operation. We work exclusively in the explicit normalization scheme, where each vertex is 
normalized according to diagrammatic isotopic convention. The differences between working in \emph{implicit} and \emph{explicit} normalization scheme, collectively called \emph{mixed normalization}, are treated in the recent anyonic DMRG paper.\cite{Pfeifer2015}

The Hermitian conjugate of the state $\ket{\Psi}$, written as 
\begin{equation}
 \bra{\Psi} = \sum_{\gamma} \Psi_{\gamma}^{\dagger} \bra{\gamma} ,
\end{equation}
is represented diagrammatically as in Fig.~\ref{AnyonicQuantumState}(ii) where the unlabeled fusion tree is reflected vertically and all its arrows are reversed. The coefficients of the vector are also complex-conjugated.

\subsubsection{Anyonic Matrix Operator}
In conventional quantum theory, an operator $\hat{O} : \mathbb{V} \rightarrow \mathbb{V}'$ is written in the bra-ket notation as,
\begin{equation}
 \hat{O} = \sum_{j',j} O_{j', j} \ket{j'} \bra{j} .
\end{equation}
where the indices $j$ and $j'$ enumerates basis states in $\mathbb{V}$ and $\mathbb{V}'$.  Example of such operators include Hamiltonians, density matrices, projectors, etc.

\begin{figure}
 \includegraphics[width=\columnwidth]{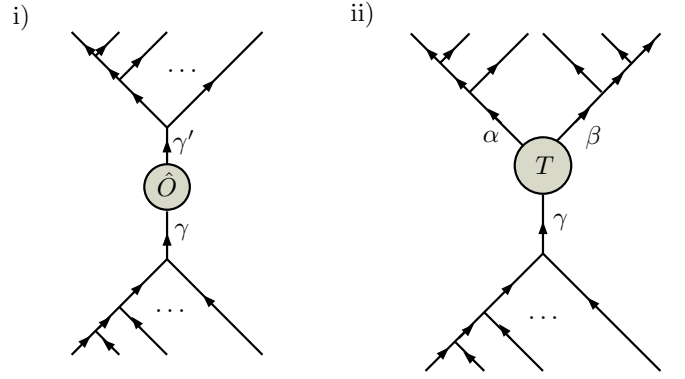}
\caption{(i) The anyonic matrix operator $\hat{O}$ and (ii) the anyonic rank-$3$ tensor $T$. The tree has been normalization according to diagrammatic isotopy. }
\label{AnyonicMatrixTensor}
\end{figure}

In a similar vein, an anyonic operator acting on a set of anyonic charges with total charge $c$ does not change the total charge.  The operator $\hat{O}_c : \mathbb{V}^{a_1, a_2, \cdots a_k}_c \rightarrow \mathbb{V}^{a'_1, a'_2, \cdots a'_{k'}}_c$ takes states of anyons $a_1, a_2, \cdots, a_k$ to states of anyons $a'_1, a'_2, \cdots, a'_{k'}$ without changing the conserved total charge $c$. As such, the operator $\hat{O}=\bigoplus_c \hat{O}_c$ can be constructed as a block-diagonal matrix with each block indexed by the conserved anyonic charge $c$. Each block matrix $\hat{O}_c$ is constructed by enumerating (as in Fig.~\ref{FusionTree}) all the fusion tree bases fusing to that charge. As such the charge-conserving matrix is indexed by the multi-index $\gamma=(c, \mu_c)$ for fusion trees and  $\gamma'=(c, \mu'_c)$ for splitting trees. The anyonic operator can therefore be written as
\begin{equation}
 \hat{O}_c = \sum_{\gamma', \gamma} \hat{O}_{\gamma', \gamma} \ket{\gamma'}\bra{\gamma} ,
\end{equation}
where $\gamma=(c, \mu_c)$ and $\gamma' = (c,\mu'_c)$ implying charge conservation. 
The matrix elements will depend on the particular physics of the system. The anyonic matrix operator is represented diagrammatically by Fig.~\ref{AnyonicMatrixTensor}(i), where the multi-indices $\gamma=(c, \mu_c)$ and $\gamma'=(c, \mu'_c)$ enumerate all the fusion and splitting trees. The vertex normalization factors of the fusion/splitting trees are absorbed into the matrix operator. 

\subsubsection{Anyonic rank-$3$ tensor}
The anyonic matrix operator can be extended to a rank-$3$ tensor where the tensor elements are indexed by three multi-indices $\alpha$, $\beta$ and $\gamma$. An anyonic rank-3 tensor $T_{\gamma}^{\alpha, \beta}$ may be represented in the manner shown in Fig.~\ref{AnyonicMatrixTensor}(ii) where the leaves on each branch of the tree are enumerated and assigned a multi-index notation $\alpha=(a, \mu_a)$, $\beta=(b, \mu_b)$ and $\gamma=(c,\mu_c)$. All the vertices on the leaves fulfill the fusion rules during enumeration of the basis, and the implicit vertex contained within the grey circle also obeys the fusion rules of the anyon model. The tensor $T_{\gamma}^{\alpha, \beta}$ is indexed by $\gamma = (c, \mu_c)$, $\alpha=(a, \mu_a)$ and $\beta=(b,\mu_b)$, where the charge triplet $(a,b,c)$, obtained from each of the subtrees, has to be compatible with the orientation of the vertex on the tensor, i.e.~$ a \times b \rightarrow c $. The explicit form of the tensor $T$ is 
\begin{equation}
 T = \bigoplus_{a,b,c}N_{ab}^{c} \sum_{\mu_a, \mu_b, \mu_c =1}^{\nu_a, \nu_b, \nu_c} \left( T_c^{a,b} \right)_{\mu_c}^{\mu_a, \mu_b} \ket{\mu_a \mu_b}\bra{\mu_c} .
\end{equation}
The direct sum implies that tensor $T$ is composed blockwise from tensors indexed by the charges of the subtrees, with  each block then being indexed by the degeneracy index of the compatible fusion trees. 

{There are more objects that can be implemented to manipulate anyonic tensor networks~\cite{Pfeifer2010a, Konig2010, Pfeifer2015}, but as the MPS is a trivalent tensor network, the anyonic tensors we have reviewed are sufficient to construct the anyonic MPS.}

\section{Anyon Model Data}
An anyon model is minimally specified by the following data: a set of charges $\mathcal{A}$, fusion rules for the charges $N_{ab}^c$, the braid matrix $R$, and the F-tensor $F$. All other quantities can be derived from these data. The anyon models we used for testing our numerical method in this paper are Fibonacci anyons, $\mathbb{Z}_2$ (spinless) fermions and $\mathbb{Z}_{\infty}$ bosons, but the charge spectrum of the physical site is restricted to $\mathcal{A}_{\text{HCB}} = \{0, 1\}$ corresponding to the hardcore constraint.  

\subsubsection{Fibonacci anyon data}
The Fibonacci anyon model consists of two charges, vacuum ($\mathbb{I}$) and Fibonacci anyon ($\tau$). 
Hence $\mathcal{A} = \{\mbb{I},\tau\}$ where the charges have quantum dimensions, $d_\mbb{I} = 1$,  $d_\tau = \frac{1+\sqrt{5}}{2}$ respectively. The fusion rules obeyed by charges are 
\begin{equation}
\mbb{I}\times \mbb{I}=\mbb{I},\quad \mbb{I} \times \tau = \tau \times \mbb{I} = \tau, \qquad  \tau \times \tau = \mbb{I} + \tau.
\end{equation}
The fusion tensor $N$ has components $N_{ab}^c = 0$ when $a \times b \nrightarrow c$ for all $a,b,c \in \mathcal{A}$. The nonzero components are given by 
\begin{equation}
N_{\mbb{I}\mbb{I}}^\mbb{I} = N_{\tau\mbb{I}}^\tau = N_{\mbb{I}\tau}^\tau = N_{\tau\tau}^\mbb{I} = N_{\tau\tau}^\tau = 1 \; . 
\end{equation}
The $R$-matrix has nonzero components
\begin{equation}
R^{\tau\tau}_\mbb{I} = e^{-4\pi i/5}, ~ R^{\tau\tau}_\tau = e^{3\pi i/5} , ~ R^{\mbb{I}\tau}_\tau = R^{\tau\mbb{I}}_\tau = R^{\mbb{I}\mbb{I}}_\mbb{I} = 1 ,
\end{equation}
for compatible charges and zero otherwise. The nontrivial $F$-move coefficients are
\begin{equation}
\left( F^{\tau\tau\tau}_\tau \right)_e^f = 
  \begin{pmatrix}
    \phi^{-1} & \phi^{-\frac{1}{2}} \\
    \phi^{-\frac{1}{2}} & -\phi^{-1} 
  \end{pmatrix} ,
\end{equation}
where $\phi = \frac{1+\sqrt{5}}{2}$, and $e,f \in \{\mbb{I},\tau\}$. The remaining $F$-move coefficients are given by
\begin{equation}
\left( F^{abc}_d \right)_e^f = N_{ab}^e N_{bc}^f N_{ec}^d N_{af}^d .
\end{equation}

\subsubsection{Fermions and Bosons Data}
Fermions and bosons can be studied within the theory of anyons and consequently  using anyonic tensor networks such as the anyonic MPS. The wave functions of fermions and bosons acquire phase factors of $-1$ and $+1$ respectively under particle pair exchange. 

The particle spectrum $\mathbb{Z}_{\infty}$ of bosons is the set of positive integer charges, denoted as $\mathbb{Z}_{\infty} = \{0, 1, 2, \ldots  \}$ with the fusion rule being ordinary addition, while the charge spectrum of fermions is $\mathbb{Z}_{2}= \{0,1 \}$ with fusion rules corresponding to addition modulo $2$, so $1 \times 1 \rightarrow 0$. 
The quantum dimensions are trivial with $d_0=d_1 = 1$ for fermions and $d_q=1~\forall~q$ for bosons. 
The fermionic exchange factors (permutation factors) are encoded in the 
fermionic $R$-matrix as
\begin{equation}
R^{00}_0 = R^{10}_1 = R^{01}_1 = 1, ~~ R^{11}_0 = -1,
\end{equation}
while for bosons all valid entries are trivially equal to one. 
The $F$-matrix for both particle types fulfills
\begin{equation}
\left( F^{abc}_d \right)_e^f = N_{ab}^e N_{bc}^f N_{ec}^d N_{af}^d .
\end{equation}

\section{Derivation of the anyonic $t$-$J$ Hamiltonian on a ladder}
We now give the explicit derivation of the Hamiltonian $\hat{h}$ on a plaquette. This Hamiltonian consists of itinerant and (Heisenberg) interaction terms written as
\begin{align}
 \hat{h} &= -t_h \left( \hat{b}_{a \rightarrow c=\mbb{I}}  + \hat{b}_{b \rightarrow d=\mbb{I}} + \text{h$\cdot$c} \right)  + J_h \left( \hat{\Pi}^{\mathbb{I}}_{a,c} + \hat{\Pi}^{\mathbb{I}}_{b,d} \right) \nonumber \\
& \quad - \frac{t_v}{2} \left( \hat{b}_{a \rightarrow b=\mbb{I}} + \hat{b}_{c \rightarrow d=\mbb{I}} + \text{h$\cdot$c} \right) + \frac{J_v}{2} \left( \hat{\Pi}^{\mathbb{I}}_{a,b} + \hat{\Pi}^{\mathbb{I}}_{c,d} \right) .
\end{align}
We derive the matrix representation of each term of the Hamiltonian in the first basis $\{\ket{I} \}$ shown in Fig.~\ref{FusionOrderPlaquette}(i). For operators like $\hat{b}_{a \rightarrow c=\mbb{I}}$, $\Pi_{ac}^0$, etc., which couple anyons on the legs of the plaquette, we transform the basis $\ket{I}$ to the basis $\ket{II}$ in Fig.~\ref{FusionOrderPlaquette}(ii), derive the action of the Hamiltonian in the basis $\ket{II}$, then transform back to the basis $\ket{I}$. 

Therefore we first show how the two bases transform, and later show the derivation of the Hamiltonian for the plaquette. 

\subsection{Fusion Tree Basis Transformation}\label{FusionBasisTransform}
The transformation between the two chosen bases of Fig.~\ref{FusionOrderPlaquette} are obtained as follows:
\begin{widetext}
\begin{equation}
 \includegraphics{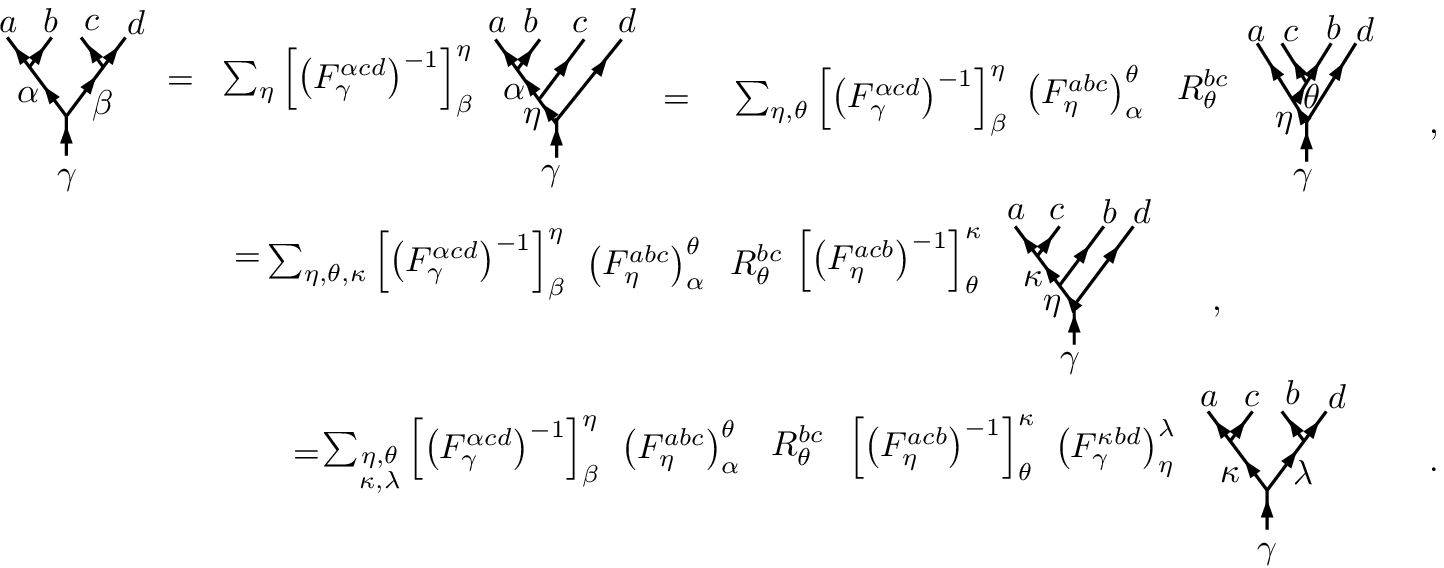}\label{A2}
\end{equation}
\end{widetext}

The above equation can be written more succinctly as
\begin{equation}
\includegraphics{FusionBasisTransform} 
\end{equation}
where the tensor $Q_{\alpha, \beta}^{\kappa, \lambda}$ is defined according to
\begin{equation}
\begin{array}{lll}
Q_{\alpha, \beta}^{\kappa, \lambda} &=& \displaystyle{\sum_{\substack{\eta, \theta \\ \kappa, \lambda}}} \left[\left( F^{\alpha cd}_{\gamma} \right)^{-1}\right]^{\eta}_{\beta} \left( F^{a b c}_{\eta} \right)^{\theta}_{\alpha} R^{bc}_{\theta} \left[\left( F^{a c b}_{\eta } \right)^{-1}\right]^{\kappa}_{\theta}\\
&&\times  \left( F^{\kappa b d}_{\gamma} \right)^{\lambda}_{\eta} ~.
\end{array}
\end{equation}
Using Dirac bra-ket notation, Eq.~\ref{A2} can alternatively be written as
\begin{equation}\label{A3}
\ket{(ab;\alpha)(cd;\beta)(\alpha \beta; \gamma)} = \sum_{\kappa, \lambda} 
Q_{\alpha \beta}^{\kappa \lambda} \ket{(ac;\kappa)(bd;\lambda)(\kappa \lambda; \gamma)} .
\end{equation}

\subsection{Anyonic $t$-$J$ Hamiltonian on a plaquette} \label{HamiltonianAnyonicLadder}
The anyonic local Hamiltonian $\hat{h}$ on a plaquette is given by
\begin{align}
 \hat{h} &= -t_h \left( \hat{b}_{a \rightarrow c=\mbb{I}}  + \hat{b}_{b \rightarrow d=\mbb{I}} + \text{h$\cdot$c} \right)  - J_h \left( \hat{\Pi}^{\mathbb{I}}_{a,c} + \hat{\Pi}^{\mathbb{I}}_{b,d} \right) \nonumber \\
& \quad - \frac{t_v}{2} \left( \hat{b}_{a \rightarrow b=\mbb{I}} + \hat{b}_{c \rightarrow d=\mbb{I}} + \text{h$\cdot$c} \right) -  \frac{J_v}{2} \left( \hat{\Pi}^{\mathbb{I}}_{a,b} + \hat{\Pi}^{\mathbb{I}}_{c,d} \right) .
\end{align}
Whereas charge label $a$ may take any value from the particle spectrum, the vacuum charge will be denoted by $\mbb{I}$ and a nontrivial anyonic charge by $a_0$. For example, in the Fibonacci anyon theory, $a_0 = \tau$. The derivation is quite general and can be used with any anyon model. Note that numerical factors such as vertex normalization factors and loop factors are not accounted for here. We account for them during the implementation of the anyonic TEBD algorithm. 

We proceed by first deriving all the kinetic energy terms and then derive all the interaction terms similarly. All the operators in the Hamiltonian are applied to the fusion tree on the left hand of Eq.~\ref{A2} which is represented in Dirac notation in Eq.~\ref{A3}. 

\underline{Kinetic terms}:
The terms contributing to the kinetic energy are the braid operators, whose matrix elements are derived below. 

(i) The matrix element of the braid operator $\hat{b}_{a \rightarrow b=\mbb{I}}$ is given by
\begin{equation}
\langle \hat{b}_{a \rightarrow b=\mbb{I}} \rangle = \delta_{a,a_0} \delta_{b,\mbb{I}}, \delta_{a',b} \delta_{b',a} \delta_{c',c} \delta_{d',d} \delta_{{\alpha}',{\alpha}} \delta_{{\beta}',\beta}   ,
\end{equation}
where we have used, for the sake of conciseness, the notation $\langle \hat{b}_{a \rightarrow b=\mbb{I}} \rangle$ as a shorthand for
\begin{equation}
\langle (a'b';\alpha')(c'd';\beta')(\alpha' \beta'; \gamma) \hat{b}_{a \rightarrow b=\mbb{I}} (ab;\alpha)(cd;\beta)(\alpha \beta; \gamma) \rangle .
\end{equation}

(ii) The matrix element of the braid operator $\hat{b}_{c \rightarrow d=\mbb{I}}$ is given by
\begin{equation}
\langle \hat{b}_{c \rightarrow d=\mbb{I}} \rangle = \delta_{c,c_0} \delta_{d,\mbb{I}}, \delta_{a',a} \delta_{b',b} \delta_{c',d} \delta_{d',c} \delta_{{\alpha}',{\alpha}} \delta_{{\beta}',\beta}    .
\end{equation}

(iii) The matrix element of the operator $\hat{b}_{a \rightarrow c=\mbb{I}}$ involves braiding of anyonic charge $a$ with $b$.  The charge $c$ has to be vacuum  for the process to have a nonzero amplitude. Its action on the basis $ \ket{(ab;\alpha)(cd;\beta)(\alpha \beta; \gamma)}$ is given by
\begin{equation}
\begin{array}{lll}
&&\hat{b}_{a \rightarrow c=\mbb{I}} \ket{(ab;\alpha)(cd;\beta)(\alpha \beta; \gamma)} \\
&& = \sum_{\kappa, \lambda} Q_{\alpha \beta}^{\kappa \lambda} \hat{b}_{a \rightarrow c=\mbb{I}} \ket{(ac;\kappa)(bd;\lambda)(\kappa \lambda; \gamma)} , \nonumber \\
& &= \sum_{\kappa, \lambda} Q_{\alpha \beta}^{\kappa \lambda} \delta_{a,a_0} \delta_{c,\mbb{I}} \ket{(ca;\kappa)(bd;\lambda)(\kappa \lambda; \gamma)} .
\end{array}
\end{equation}
The expectation value $\langle \hat{b}_{a \rightarrow c=\mbb{I}} \rangle$ is
\begin{equation}
\begin{array}{lll}
\langle \hat{b}_{a \rightarrow c=\mbb{I}} \rangle &=& \sum_{\substack{\kappa',{\lambda}' \\ \kappa, \lambda}} \bra{(a'c';{\kappa}')(b'd';{\lambda}')({\kappa}' {\lambda}'; \gamma)} Q_{\alpha' \beta'}^{* \kappa' \lambda'} Q_{\alpha \beta}^{\kappa \lambda}\\
&&\times  \delta_{a,a_0} \delta_{c,\mbb{I}} \ket{(ca;\kappa)(bd;\lambda)(\kappa \lambda; \gamma)} ,
\end{array}
\end{equation}
which simplifies to
\begin{equation}
\langle \hat{b}_{a \rightarrow c=\mbb{I}} \rangle = \sum_{\kappa, \lambda} Q_{\alpha \beta}^{\kappa \lambda} (Q^{\dagger})^{\alpha' \beta'}_{\kappa \lambda} \delta_{a,a_0} \delta_{c,\mbb{I}} \delta_{a',c} \delta_{c',a} \delta_{b',b} \delta_{d',d} .
\end{equation}

(iv) The expectation value $\langle \hat{b}_{b \rightarrow d=\mbb{I}} \rangle$ is similarly given by
\begin{equation}
\langle \hat{b}_{b \rightarrow d=\mbb{I}} \rangle = \sum_{\kappa, \lambda} Q_{\alpha \beta}^{\kappa \lambda} (Q^{\dagger})^{\alpha' \beta'}_{\kappa \lambda} \delta_{b,b_0} \delta_{d,\mbb{I}} \delta_{a',a} \delta_{c',c} \delta_{b',d} \delta_{d',b} .
\end{equation}

\underline{Interaction terms}:
The interaction terms consist of projectors whose matrix elements are derived similarly to the braid terms. The projection favours fusion of nontrivial anyons to the vacuum charge.

i) The action of the projector $\hat{\Pi}_{a,b}^{\mbb{I}}$ on the fusion basis is given as
\begin{equation}
\hat{\Pi}_{ab}^{\mbb{I}} \ket{(ab; \alpha)(cd;\beta)(\alpha \beta;\gamma)}
= \Pi_{ab}^{\alpha} \ket{(ab; \alpha)(cd;\beta)(\alpha \beta;\gamma)}
\end{equation}
where the element $\Pi_{ab}^{\alpha} = 1$ if $\alpha = \mbb{I}$ (vacuum) and $a=a_0$, $b=b_0$ (i.e.~nontrivial charges). The expectation value of the projector $\hat{\Pi}_{ab}^{\mbb{I}}$ is thus
\begin{equation}
 \langle \hat{\Pi}_{ab}^{\alpha=\mbb{I}} \rangle = \delta_{a',a_0} \delta_{b',b_0} \delta_{a',a} \delta_{b',b} \delta_{\alpha',\alpha} \delta_{\alpha,\mbb{I}} \delta_{c',c} \delta_{d',d} \delta_{\beta,\beta'} .
\end{equation}

(ii) The matrix element $\hat{\Pi}_{cd}^{\mbb{I}}$ of the projector is similarly given as
\begin{equation}
 \langle \hat{\Pi}_{cd}^{\beta=\mbb{I}} \rangle = \delta_{a',a} \delta_{b',b} \delta_{\alpha',\alpha} \delta_{c,c_0} \delta_{d,d_0} \delta_{\beta,0} \delta_{c',c} \delta_{d',d} \delta_{\beta,\beta'} .
\end{equation}

(iii) The action of the projector $\hat{\Pi}_{ac}^{\mbb{I}}$ on the basis $\ket{(ab;\alpha)(cd;\beta)(\alpha \beta; \gamma)}$ is 
\[
\begin{array}{lll}
&\hat{\Pi}_{ac}^{\mbb{I}}& \ket{(ab;\alpha)(cd;\beta)(\alpha \beta; \gamma)} \\
&&= \displaystyle{\sum_{\kappa, \lambda} Q_{\alpha \beta}^{\kappa \lambda} \Pi_{ac}^{\kappa} \ket{(ac;\kappa)(bd;\lambda)(\kappa \lambda; \gamma)} } \\
&&= \sum_{\kappa, \lambda} Q_{\alpha \beta}^{\kappa \lambda} \delta_{a,a_0} \delta_{c,c_0} \delta_{\kappa,\mbb{I}} \ket{(ac;\kappa)(bd;\lambda)(\kappa \lambda; \gamma)} .
\end{array}
\]
The matrix element $\langle \hat{\Pi}_{ac}^{\mbb{I}} \rangle$ is 
\begin{equation}
\langle \hat{\Pi}_{ac}^{\mbb{I}} \rangle = \sum_{\kappa, \lambda} Q_{\alpha \beta}^{\kappa \lambda} (Q^{\dagger})^{\alpha' \beta'}_{\kappa \lambda} \delta_{\kappa, \mbb{I}} \delta_{a,a_0} \delta_{c,c_0} \delta_{a',a} \delta_{c',c} \delta_{b',b} \delta_{d',d} .
\end{equation}

(iv) The matrix element $\langle \hat{\Pi}_{bd}^{\mbb{I}} \rangle$ is similarly given by
\begin{equation}
\langle \hat{\Pi}_{bd}^{\mbb{I}} \rangle = \sum_{\kappa, \lambda} Q_{\alpha \beta}^{\kappa \lambda} (Q^{\dagger})^{\alpha' \beta'}_{\kappa \lambda} \delta_{b,b_0} \delta_{d,d_0} \delta_{\lambda,\mbb{I}} \delta_{a',a} \delta_{b',b} \delta_{c',c} \delta_{d',d} .
\end{equation}

\end{document}